\documentclass[preprint,journal]{vgtc}            

\onlineid{0}

\vgtccategory{Research}

\vgtcpapertype{application/design study}

\title{
Topographic Visualization of Near-surface Temperatures\\
for Improved Lapse Rate Estimation\\ 
}

\author{%
  \authororcid{Kevin Höhlein}{0000-0002-4483-8388},
  \authororcid{Timothy Hewson}{0000-0002-3266-8828}, and 
  \authororcid{Rüdiger Westermann}{0000-0002-3394-0731}
}

\authorfooter{
  \item
  	Kevin Höhlein is with Technical University of Munich. \\
  	E-mail: kevin.hoehlein@tum.de.
  \item
  	Timothy Hewson is with the European center for medium-range weather forecasting (ECMWF).
  	E-mail: Timothy.Hewson@ecmwf.int.

  \item Rüdiger Westermann is with Technical University of Munich.\\
  	E-mail: westermann@tum.de.
}

\abstract{%
    Numerical model forecasts of near-surface temperatures are prone to error. This is because terrain can exert a strong influence on temperature that is not captured in numerical weather models due to spatial resolution limitations. To account for the terrain height difference between the forecast model and reality, temperatures are commonly corrected using a vertical adjustment based on a fixed lapse rate. This, however, ignores the fact that true lapse rates vary from $1.2\,\text{K}$ temperature drop per $100\,\text{m}$ of ascent to more than $10\,\text{K}$ temperature rise over the same vertical distance. In this work, we develop topographic visualization techniques to assess the resulting uncertainties in near-surface temperatures and reveal relationships between those uncertainties, features in the resolved and unresolved topography, and the temperature distribution in the near-surface atmosphere. Our techniques highlight common limitations of the current lapse rate scheme and hint at their topographic dependencies in the context of the prevailing weather conditions. 
    Together with scientists working in postprocessing and downscaling of numerical model output, we use these findings to develop an improved lapse rate scheme. This model adapts to both the topography and the current weather situation. We examine the quality and physical consistency of the new estimates by comparing them with station observations around the world and by including visual representations of radiation-slope interactions.
}


\graphicspath{{figs/}{figures/}{pictures/}{images/}{./}} 
\usepackage[dvipsnames,svgnames]{xcolor}
\usepackage{tabu}                      
\usepackage{booktabs}                  
\usepackage{lipsum}                    
\usepackage{mwe}                       
\usepackage{mathptmx}                  
\usepackage{amsmath}
\usepackage{amssymb}

\usepackage{todonotes}

\begin{document}

\firstsection{Introduction}

\maketitle

\begin{figure}[t]
    \centering
    \begin{tikzpicture}
        \node[inner sep=0pt] (lowres) at (0, 0) {
        \includegraphics[width=.24\textwidth]{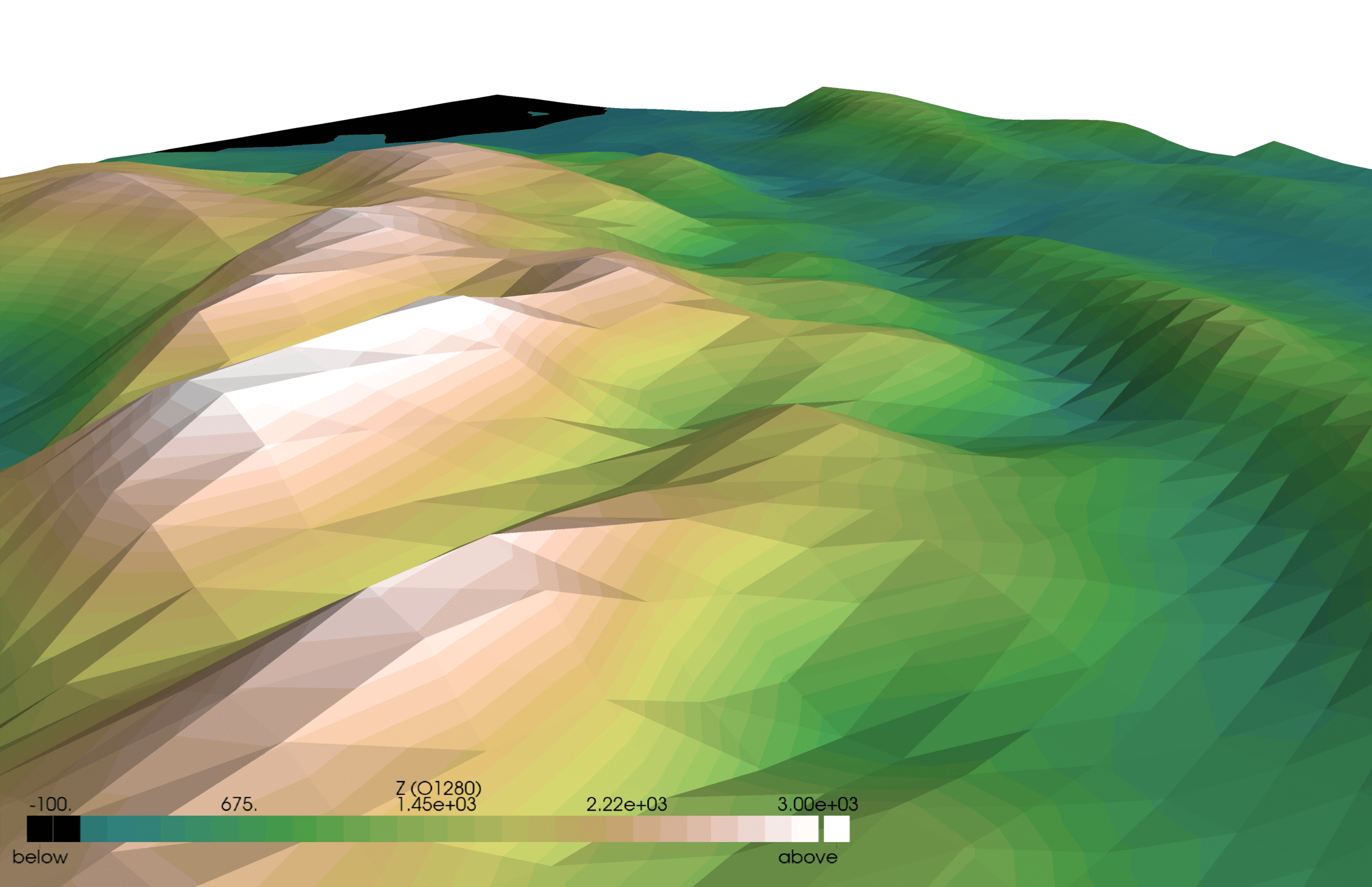}
        };
        \node at (-1.8, 1.2) {(a)};
    \end{tikzpicture}
    \begin{tikzpicture}
        \node[inner sep=0pt] (highres) at (0, 0) {
        \includegraphics[width=.24\textwidth]{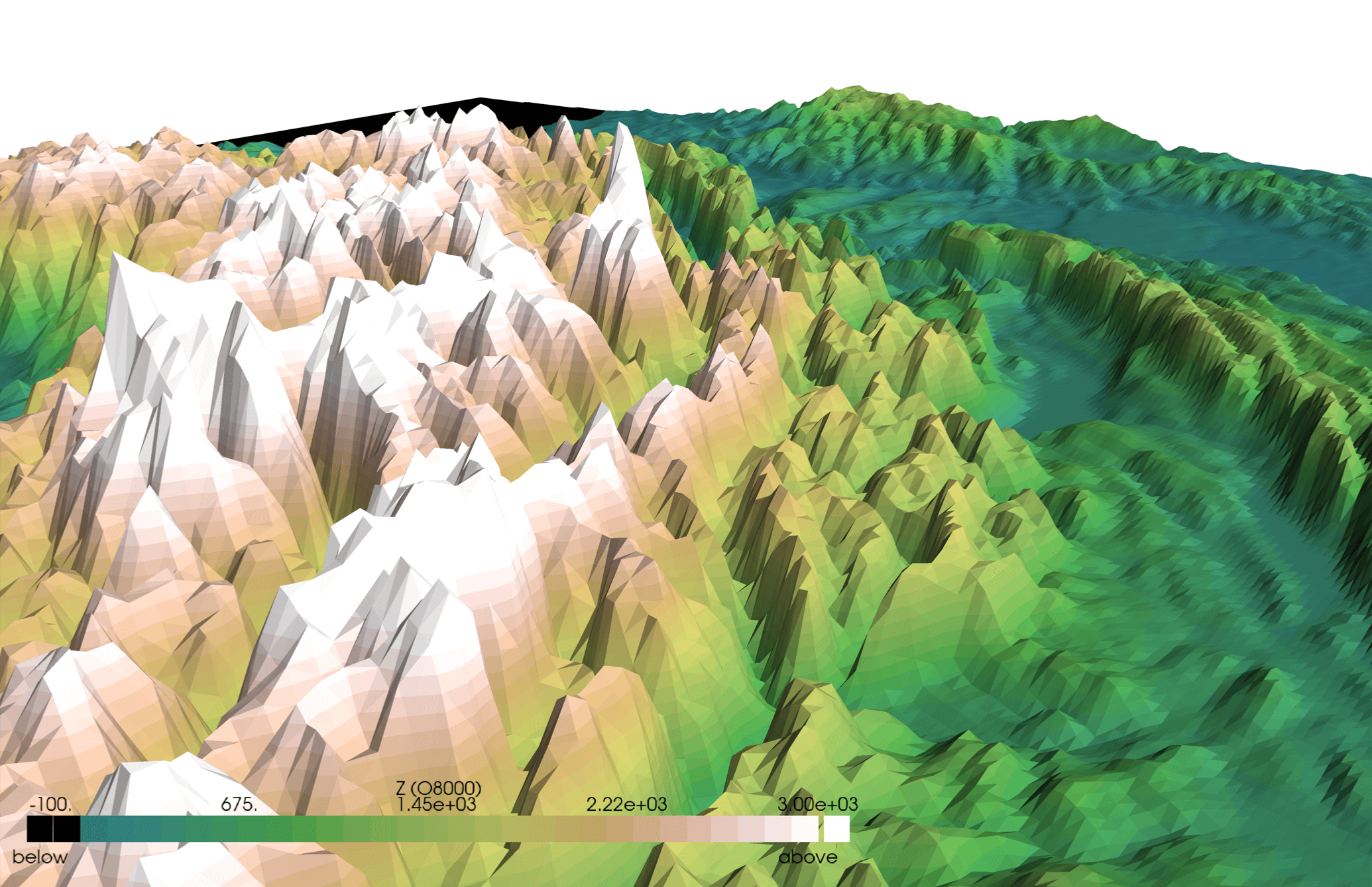}
        };
        \node at (-1.8, 1.2) {(b)};
    \end{tikzpicture} 
    \caption{Comparison of the orography around Mont Blanc at different resolutions. (a) Average grid spacing 9 km, as used in the ECMWF medium range model; (b) Average spacing 1 km.}
    \label{fig:orography}
\end{figure}

One of the most important parameters for weather forecast users is temperature. Ordinarily, and by convention, this means "2 m temperature" -- i.e., measured $\sim$ 2 m above ground. Numerical weather prediction models are generally good at forecasting 2 m temperatures over flat terrain but can struggle elsewhere. This is because 2 m temperature depends strongly on altitude and because (away from plains) the altitude of a selected site does not generally equal the altitude of the corresponding numerical model region. Numerical models have a finite spatial resolution, and within a grid box, all terrain is implicitly at the same height. For instance, the global model of the European Centre for Medium-range Weather Forecasts (ECMWF) operates on grids with a box size of approximately 9 km by 9 km, which is far too coarse to resolve topographic details. \autoref{fig:orography} illustrates this by comparing terrain representations of the same geographical region at different resolutions. Shown is the terrain around Mont Blanc, as an example, at 9 km and 1 km resolution. Topographic extremes are smoothed out significantly or dismissed entirely in the coarser representation. 

\emph{Downscaling methods} are needed to correct the model outputs and remove the low-res bias. An introduction to the goals and principles of basic downscaling methods can be found, e.g., in Wilby and Wigley~\cite{wilby1997downscaling}. To correct near-surface temperatures for terrain altitude mismatches, a common approach is modeling the local lapse rate -- i.e., the rate of temperature change with height -- and using the terrain height difference as a multiplier for this to estimate the required correction. Simple correction schemes apply a fixed lapse rate such that, e.g., temperatures always drop, going upwards, by $0.65$ K per 100 m, or $6.5$ K/km, as specified in the standard Atmosphere, defined by the International Civil Aviation Organization (ICAO) \cite{international1993manual}. 

In practice, however, lapse rates vary a lot in different situations. 
In sunny weather, near-surface air tends to be markedly warmer than air higher up. The temperature drop rate then reaches around 10 K/km before the air masses become unstable to convection. Indeed, values of 12 K/km can be reached temporarily during strong insolation. 

\begin{figure}[!h]
    \centering
    \begin{tikzpicture}
        \node[inner sep=0pt] (lowres) at (0, 0) {
        \includegraphics[width=.24\textwidth]{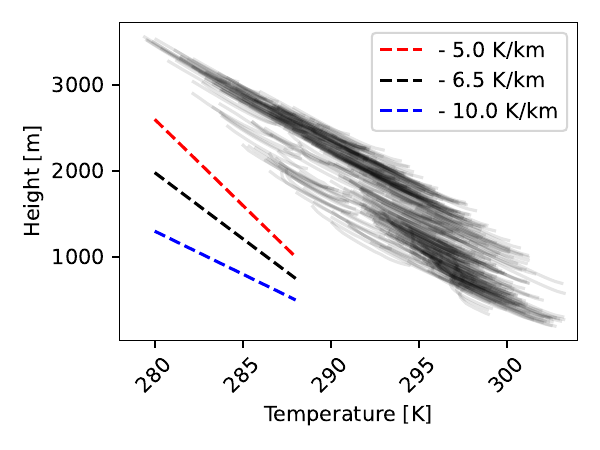}
        };
        \node at (0.2, 1.2) {(a)};
    \end{tikzpicture}
    \begin{tikzpicture}
        \node[inner sep=0pt] (lowres) at (0, 0) {
        \includegraphics[width=.24\textwidth]{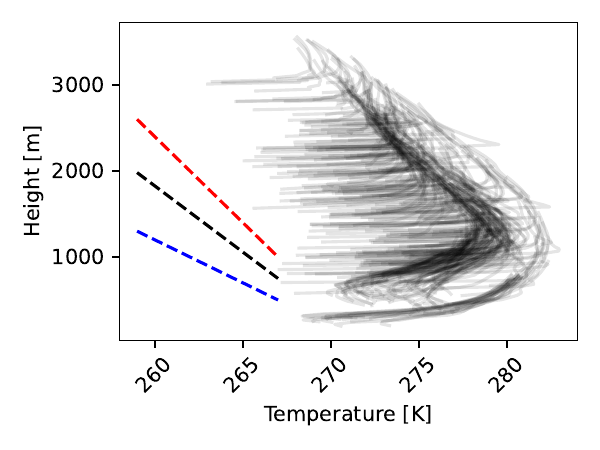}
        };
        \node at (1.6, 1.2) {(b)};
    \end{tikzpicture}
    \caption{Vertical temperature profiles in the bulk atmosphere above selected grid points in the domain of \autoref{fig:orography}. Each line represents the temperature profile over one grid point. Dashed lines are shown for reference. (a) Temperature profiles on a summer afternoon (July 23, 2021, 1400 UTC); the profiles are dominated by regular negative temperature gradients. (b) Inversion situation on a winter morning (December 19, 2021, 0600 UTC); at the lower end of the profiles, the temperature increases with altitude, indicating an inversion situation close to the earth's surface. Towards the upper end of the profiles -- i.e., away from the surface -- the regular temperature stratification is resumed.}
    \label{fig:profiles}
\end{figure}

In other scenarios, the drop rate can fall below 5 K/km and even reverse its sign. Weather situations where the temperature rises with increasing altitude are called inversions. In calm and clear weather conditions, inversions can lead to positive temperature gradients of, say, 100 K/km across tens or hundreds of meters.  Diurnal variations in lapse rate are also commonplace, especially in light wind situations with predominantly clear skies.
\autoref{fig:profiles} illustrates temperature profiles associated with regular and more unusual weather conditions. 


In this study, we set out to improve upon the fixed lapse rate assumption by using 3D topographic visualizations to assist in understanding forecast-observation mismatches and to evaluate and optimize an alternative adjustment approach based on dynamically varying lapse rates. Through analysis of the immediate impact of adjustments to lapse rate parameters on model temperature output, one can clarify how, when, and where there is a high sensitivity, obtain enhanced process understanding, and eventually improve the predictions. On the one hand, this requires a methodology to effectively locate sensitive behavior precisely in space, but on the other, there is a need to contextualize relative to the geospatial frame and simulated atmospheric processes. To achieve both, we present topographic visualizations to assess the sensitive behavior of estimated 2 m temperature to lapse rate definitions and reveal relationships between sensitive regions, features in low- and high-res orography, and the vertical temperature distribution in the near-surface atmosphere. 

\subsection*{Contribution}

We propose an interactive visualization workflow to
\begin{itemize}
    \item picture how downscaled 2 m temperature varies with height,
    \item identify common limitations of the current lapse rate scheme, including its topographic dependencies in the context of the prevailing weather conditions,
    \item compare low-res and station site near-surface temperatures and visualize temperature distribution in the near-surface atmosphere,
    \item assess the downscaling accuracy by comparing with observations. 
\end{itemize}

We use the proposed workflow for analyzing near-surface temperature, downscaled from model data, i.e., orography and hourly temperature fields, at (e.g.) 9 km spatial resolution to 1km. Our work is motivated by the following analysis questions relevant to different user groups in meteorology: 

\begin{itemize}

\item To what degree do 2 m temperature values corrected with a fixed lapse rate assumption make physical sense in different topographic/meteorological settings? (\textbf{Q1})

\item When fixed-lapse-rate-based correction fails, how do the resulting errors relate to temperature distribution in the near-surface atmosphere? (\textbf{Q2})

\item What is the impact of changing the lapse rate formula, and how well can such alternatives correct the 2 m temperature? (\textbf{Q3}) 

\item What is the accuracy of 2 m temperature fields, corrected in different ways, relative to independent surface station observations, scattered across the terrain and expected to be imperfect due to measurement and metadata errors in different classes? (\textbf{Q4})
\end{itemize}

\textbf{Q1} to \textbf{Q4} are relevant to scientists working in postprocessing and downscaling of numerical model output, as well as forecasters and specialist forecast users where the emphasis shifts more to real-world scenarios such as, e.g., operating a ski resort. The code for the project is publicly available \cite{hoehlein2024code}

\section{Background and related works}

Our dynamic lapse rate scheme was motivated by work by Sheridan et~al.~\cite{sheridan_main,sheridan_2}. These studies focused on small regions of the British Isles, using data from a limited area model as input. Here, we expand to use a global domain instead. In Sheridan et~al.~\cite{sheridan_main}, the input data spatial resolution ($L$) $ = 4\ \text{km}$, whilst in our case $L = 9 \ \text{km}$. Our data is sourced from the operational global medium-range weather forecasts produced by the ECMWF, where one author is based. This step down in resolution presents some different challenges, although the overarching scientific hypothesis we use is similar. 

Regarding output data, Sheridan et~al.~\cite{sheridan_main} present values at a set of measurement sites in a region of England; here, our aim is instead to expand substantially by devising methods that are scalable to very-high-resolution \textit{global} grids (1 km in this study, see \autoref{sec:data}), whilst still using high-density observations for verification. Sheridan et~al.~\cite{sheridan_2} describe some complex refinements to the earlier study, such as the representation of cold air pooling in valleys. In this study, those are not explicitly used, partly for simplicity and partly because topographic characteristics worldwide are much more diverse than in the UK.
From the application perspective, our main aim is to improve upon the standard lapse rate assumption to deliver better forecasts whilst keeping the new method explainable and intelligible for users.

\subsection{Downscaling surface temperatures} 

Given low-res model outputs and higher-res orography fields, common operational downscaling schemes use horizontal interpolation (or "nearest neighbor") in combination with a vertical correction according to a constant lapse rate linked to the ICAO standard atmosphere \cite{international1993manual}. Especially in regions with complex orography, this simplistic approach leads to physically implausible or inaccurate predictions, which must undergo further postprocessing to produce useful predictions (cf., e.g., Fiddes and Gruber~\cite{fiddes2014toposcale}). 

More elaborate downscaling and interpolation approaches have been developed (e.g., \cite{frei2014interpolation,hiebl2016daily,luo2019assessment,lussana2018three}) but often come with a significant compute footprint (as in full dynamical downscaling -- e.g., \ two of the three methods in\cite{kruyt2022downscaling}) or data requirements and are therefore difficult to deploy to global-scale applications. Several studies rely on observations (rather than model data) as input and derive a range of complex techniques for handling those \cite{frei2014interpolation,hiebl2016daily,lussana2018three}. Whilst these engender more vertical lapse rate complexity and potentially greater accuracy, the techniques are considered too involved for global application. Also, the needed high-density observations are missing in most parts of the world. Numerical methods for downscaling meteorological variables to sub-grid resolution use orography-related predictors at high spatial resolution \cite{fiddes2012toposub}, downscale temperatures by interpolating pressure-level data \cite{fiddes2014toposcale}, or quantile mapping is applied in postprocessing to compensate for station-wise statistical biases \cite{fiddes2022topoclim}. The authors of \cite{frei2014interpolation} suggest fitting nonlinear vertical temperature profiles with compact parametric forms emulating the vertical variation in temperature. The approach uses a two-step procedure by first estimating a background temperature field on coarse spatial resolution, which is then superimposed with the vertical variation. So-called optimal interpolation methods have been applied to derive high-res temperature maps from high-res observation networks in the Alps region \cite{uboldi2008three} and Norway \cite{lussana2018three}.
The authors of \cite{luo2019assessment} propose to reduce elevation-related biases in reanalysis datasets using an elevation correction method with internal lapse rates derived from different reanalysis pressure levels.

\subsection{Meteorological map visualisation}

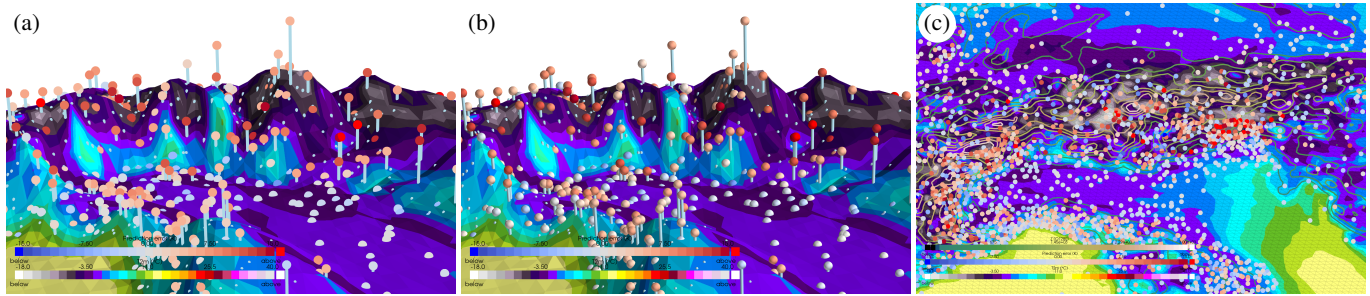
\begin{figure*}
    \centering
    \begin{tikzpicture}
        \node[inner sep=0pt, anchor=north west] (lowres) at (0, 0) {
        \includegraphics[width=.33\textwidth]{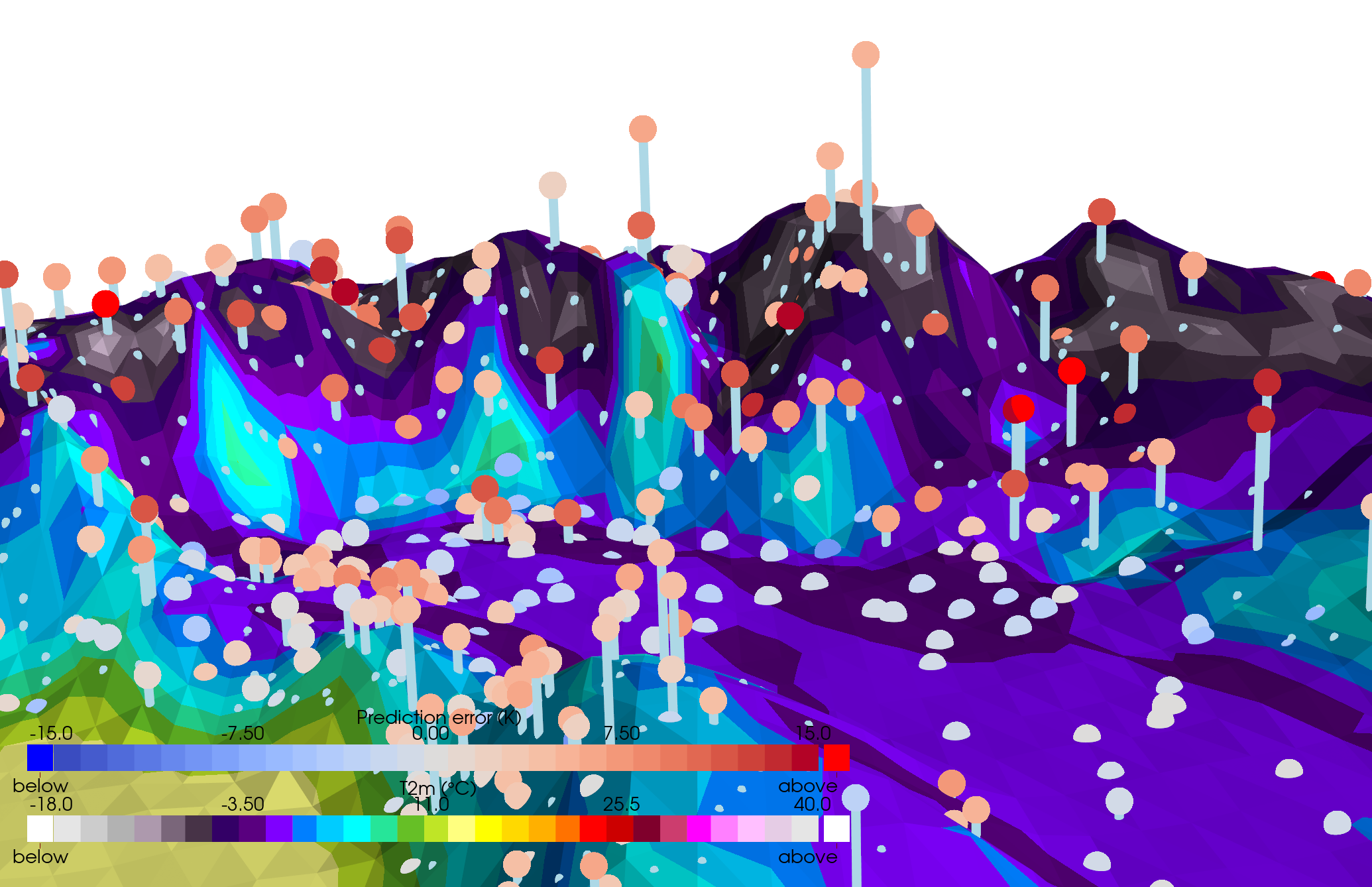}
        };
        \node[circle, inner sep=0pt, anchor=north west, fill=white, minimum size = 5mm] at (0.1, -0.1) {(a)};
    \end{tikzpicture}
    \begin{tikzpicture}
        \node[inner sep=0pt, anchor=north west] (lowres) at (0, 0) {
        \includegraphics[width=.33\textwidth]{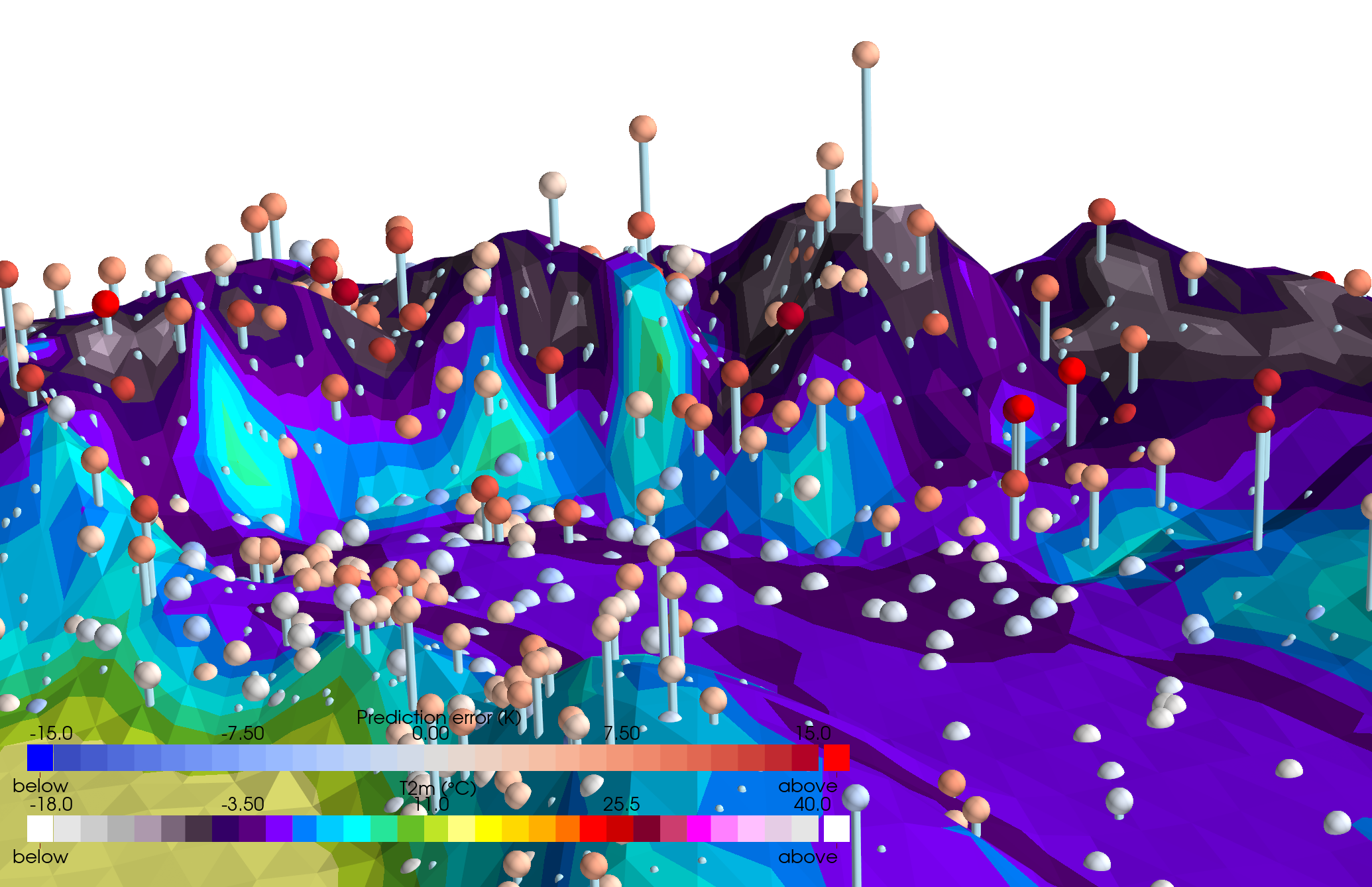}
        };
        \node[circle, inner sep=0pt, anchor=north west, fill=white, minimum size = 5mm] at (0.1, -0.1) {(b)};
    \end{tikzpicture}
    \begin{tikzpicture}
        \node[inner sep=0pt, anchor=north west] (lowres) at (0, 0) {
        \includegraphics[width=.33\textwidth]{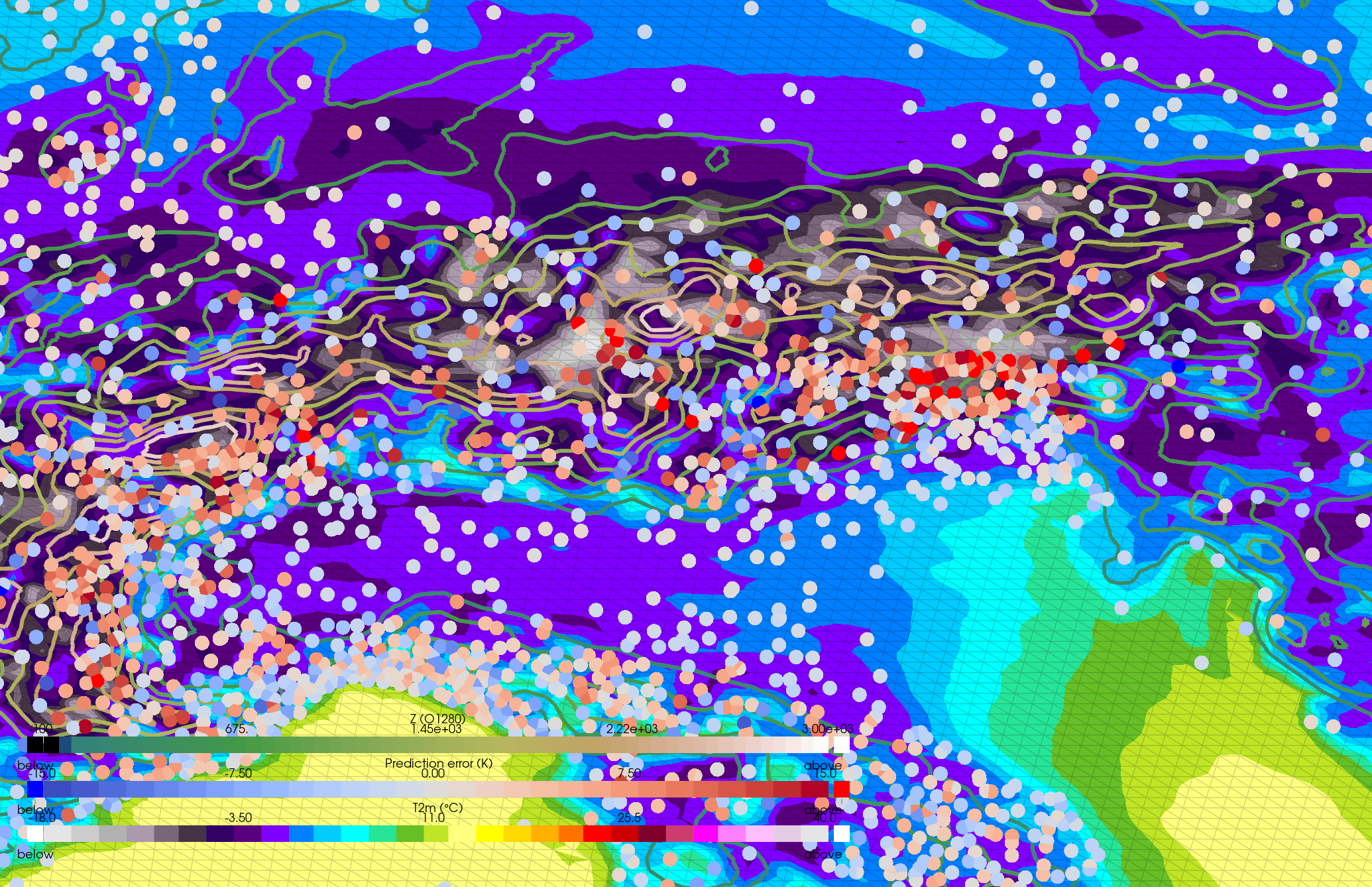}
        };
        \node[circle, inner sep=0pt, anchor=north west, fill=white, minimum size = 5mm] at (0.1, -0.1) {(c)};
    \end{tikzpicture}
    \caption{Low-res orography with near-surface temperature encoded in color. Station locations are shown as spheres, with their color displaying temperature differences. Vertical lines below station sites provide a reference for locating the stations in the 2D domain. Dots on the terrain suggest the presence of stations below the surface at this location. The shading of the spheres can be toggled off (a) or on (b) to enhance the readability of the color scale. A 2D map-like view (c) is obtained by toggling a bird view with parallel projection along the elevation axis.
    }
    \label{fig:maps-temp}
\end{figure*}

Analyzing the spatio-temporal distribution and relations between atmospheric variables, measured and numerically simulated, is at the core of meteorological data visualization. Central to this task is the use of visualizations that can simultaneously provide views of topographic information to reveal geospatial information, such as station locations and terrain as spatial frames of reference \cite{andrienko2006exploratory}, and the physical variables on the terrain and in the surrounding atmosphere. 

The process of meteorological map-making was discussed by Monmonier \cite{Monmonier_2001}, and the review by Stephens et~al.~\cite{stephens2012communicating} focuses on probabilistic information communication in atmospheric sciences. In several follow-up summary reports, the tools and techniques in climate and weather research have been reviewed \cite{Middleton-2005,Noecke-2008}. More recently, Rautenhaus et~al.~\cite{RautenhausEtAl2018VisMet}, Aftal et~al.~\cite{Aftal-2019}, and Roeber et~al.~\cite{rober2021visualization} have provided overviews of atmospheric data visualization, including taxonomies of techniques, discussions of differences between operational use and research, as well as specific approaches in climate science. 

Advice on the generation of meteorological maps to enable effective human comprehension of the displayed data is given in the book by Hoffmann et~al.~\cite{Hoffmann-2017}. In operational settings, 2D surface maps using color coding of temperature in combination with contour lines of surface pressure are still most often used. These maps are augmented by glyphs to indicate station data and linked to domain-specific diagrams. Pressure level charts often visualize the 500 hPa level via 2D maps to represent atmospheric flow at the mid-troposphere. In this context, especially the effectiveness of visual attributes such as color has been studied \cite{Teuling-2011,Stauffer-2015}. Improved readability and better communication of quantitative meteorological variables have been reported with the perceptional linear hue-chroma-luminance color space.  An evaluation of maps and additional climate-specific visualization was pursued by Dasgupta et~al.~\cite{Dasgupta-2015}, who provide a list of design guidelines for color and visual saliency. Duebel et~al.~\cite{Duebel-2017} discuss the visualization of geospatial data on 2D height fields, i.e., terrain fields, and provide means to visually communicate simultaneously the terrain field and data, including data-associated uncertainty. 

Driven by the use of numerical ensemble simulations in atmospheric science, the visualization of uncertainty has become a major research area in recent years. Prevalent to many of the existing ensemble visualization techniques is the question of visually conveying the ensemble spread of atmospheric variables from different numerical simulations. Guidance on the mapping of uncertain variables was provided by Kaye et~al.~\cite{Kaye-2012} and Retchless and Brewer~\cite{Retchless-2016}, for instance, to combine color and pattern for visualizing climate change parameters with uncertainty. The survey by MacEachron \cite{MacEachren-2005} focuses explicitly on uncertainty in geospatial science and cartography. Griethe et~al.~\cite{griethe2006visualization} categorize uncertainty visualization into intrinsic and extrinsic techniques, depending on whether existing graphical representations are modified to convey the uncertainty or additional graphical primitives are added. Several summaries shed light on the sources and models of uncertainty \cite{Potter-2012,Bonneau-2014, wang2018, Kamal2021RecentAA}, including categorizations of uncertainty visualization techniques depending on whether stochastic uncertainty models or ensembles are used. Representative examples of atmospheric data visualizations, to name just a few, are statistical summaries \cite{PotterEtAl2010SummaryStat,codda2019}, spaghetti plots \cite{sanyal2010noodles}, contour box plots \cite{ContourBoxplotsWhitaker} and streamline variability plots \cite{Ferstl15UncEns}. Most similar to our method for visualizing confidence information across the terrain are visualizations of the effect of uncertainty on the position and structure of isosurfaces, e.g., \ by using surface displacements \cite{Grigoryan2004PointbasedPS} and confidence surfaces \cite{zehner-2010,pothkow2010positional,Pfaffelmoser-2011}.

\section{Datasets\label{sec:data}}

The dataset for our study comprises global temperature prediction data generated by the medium-range prediction system at the ECMWF. Data are available on a (cubic octahedral) reduced Gaussian grid \cite{malardel2016new} (O1280) with global coverage and average grid spacing of 9 km. Model predictions are retrieved for 2 m temperatures and volumetric temperatures for the 20 lowest model levels. As the model operates on terrain-following hybrid levels, volumetric visualizations require the precomputation of the local geometric altitude of the respective model levels. While the accurate elevation levels usually depend on the pressure and humidity distribution of the weather situation, the altitude of the lowest model levels is only marginally affected by such variations. Therefore, we approximate the model levels using a standard atmosphere assumption. This procedure does not affect the quantitative evaluation of the proposed methodology, as model-level data is not used here. We have verified that the difference between approximate and physical model levels is imperceptible for the visualizations. Hourly data are available for the time period from April 1, 2021 to March 31, 2022, resulting in a total size of the dataset of approx. 2 TB. For quantitative analysis, we use the full dataset. For visualizations, case studies are selected based on meteorological prior knowledge (see \autoref{sec:visana}). As a test case for a region with complex topographic structure, we select a geographic region between $43^\circ$ and $49^\circ$ latitude, as well as $4^\circ$ and $18^\circ$ longitude. The region is located in central Europe and covers the Alps mountain range and parts of northern Italy. 

In addition to the model elevation field, we use a high-res orography dataset with 1 km average grid spacing (O8000) provided by the ECMWF and composited together from the following sources: SRTM30 for 60S to 60N \cite{farr2007shuttle}; GLOBE for the north of 60N \cite{hastings1999global}, RAMP2 for the south of 60S \cite{radarsat}, BPRC for Greenland \cite{greenlanddem}, IS 50V for Iceland \cite{icelanddem}. A high-resolution land-sea mask is computed by downsampling a  watermask at $100\,\text{m}$ resolution \cite{mikelsons2021global} to the O8000 grid.

Global near-surface temperature observations are retrieved from the HDOBS database of the ECMWF, comprising 86 Mio.\ records of surface temperature, station location, and station elevation from more than 16000 weather stations worldwide (see \cite{ecobs}). Missing values and faulty observations reduce the number of valid records to 65 Mio.\ observations at 14500 station sites. Data records cover the time between April 1, 2021, and March 31, 2022, and are generally available multiple times a day, with frequency depending on each station's schedule. 

\section{Methods}

The topographic visualization workflow by which we address the requirements from meteorology features three different visualization options: the terrain map, including station data, the atmosphere layer, and the elevation variability plot. Additionally, control panels facilitate data selection and interaction with data processing and display. The result of applying our improved lapse rate scheme for temperature correction can be compared directly to station data.  
The tool is implemented in Python, using the visualization library PyVista~\cite{sullivan2019pyvista}, providing Python bindings to the visualization toolkit (VTK) \cite{schroeder1998visualization}. The graphical user interface is based on Python bindings of Qt5. 

\subsection{Terrain map}

The terrain map panel displays the 3D terrain field augmented by intrinsic and extrinsic visual encodings \cite{griethe2006visualization} of additional information like the temperature distribution, orography, or land cover. It provides an interactive environment allowing one to switch between low- and high-res terrain, showing differences in height between them and showing differences between the low-res surface temperature and ground truth station data. Via zooming, this interaction facilitates analysis at both global and regional scales. 

\autoref{fig:maps-temp} shows visualizations of the low-res terrain map with color coding of temperature. For temperature fields, meteorology users demand color maps that are compatible with the ones used in operational forecasting. Specifically, temperature maps should clearly distinguish temperatures with a resolution of around 2 K and follow a predefined listed color map. For lapse rate-related color maps, the user prefers color schemes that respect physical prior knowledge and allow for a simple comparison against the default lapse rate of $-6.5$ K/km. We address this by employing diverging color maps with configurable opacity functions (see \autoref{sec:visana}).

Temperature is encoded intrinsically by mapping the respective fields to color. We employ an extrinsic encoding on spheres embedded into the terrain to visualize stations and temperature measurements or differences. We provide two different visualization options: 

I) Spheres can be colored with a constant color indicating temperature or difference but without applying shading. This gives the most unobscured display of temperature values, yet it makes comparison to near-surface temperature on the terrain -- whose colors are modulated by surface shading -- difficult (see \autoref{fig:maps-temp} (a)). Constant sphere coloring, however, is advantageous if a 2D map view is generated by looking from above the terrain using an orthographic projection and showing only unmodulated surface colors (see \autoref{fig:maps-temp} (c)). 

II) Spheres can be shaded according to the selected lighting conditions to let spheres stand out less in the visualization (see \autoref{fig:maps-temp} (b)). We use the 3-light illumination model provided by PyVista, which simulates multiple lights to realize shading without letting certain sphere parts become too dark. In either case, we use additional lines to emphasize how much above or below the terrain a station is located. Note here that while accurate station positions are available, the terrain -- regardless of whether it is the low- or high-res version -- never represents orography perfectly. Since some stations are located below the terrain, the user can use transparency for the terrain surface to let these stations shine through, change the camera position, or invert the station offset to show stations below the surface on the opposite side. The line color is chosen to stand out against the white background and diverge from the colors in the temperature color map toward the extreme temperatures. 
Optionally, the user can switch to an alternative lighting mode, which simulates parallel sunlight. This is useful for investigating factors that are potentially impacting temperature anomalies (see \autoref{sec:visana}).

\begin{figure}
    \centering
        \begin{tikzpicture}
        \node[inner sep=0pt, anchor=north west] (highres) at (0, 0) {
        \includegraphics[width=.24\textwidth]{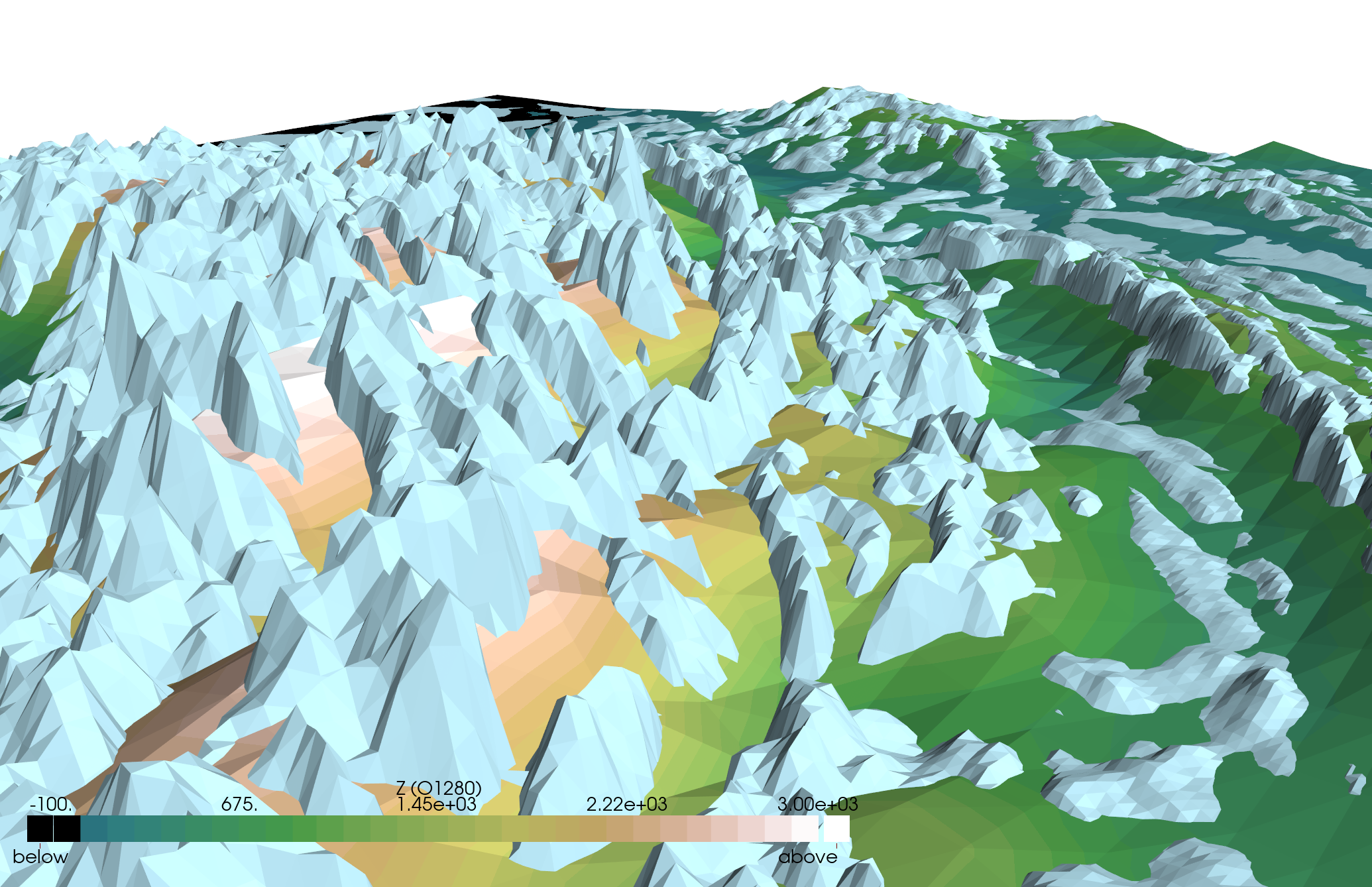}
        };
        \node[circle, inner sep=0pt, anchor=north west, fill=white, minimum size = 5mm] at (0.1, -0.1) {(a)};
    \end{tikzpicture} 
    \begin{tikzpicture}
        \node[inner sep=0pt, anchor=north west] (highres) at (0, 0) {
        \includegraphics[width=.24\textwidth]{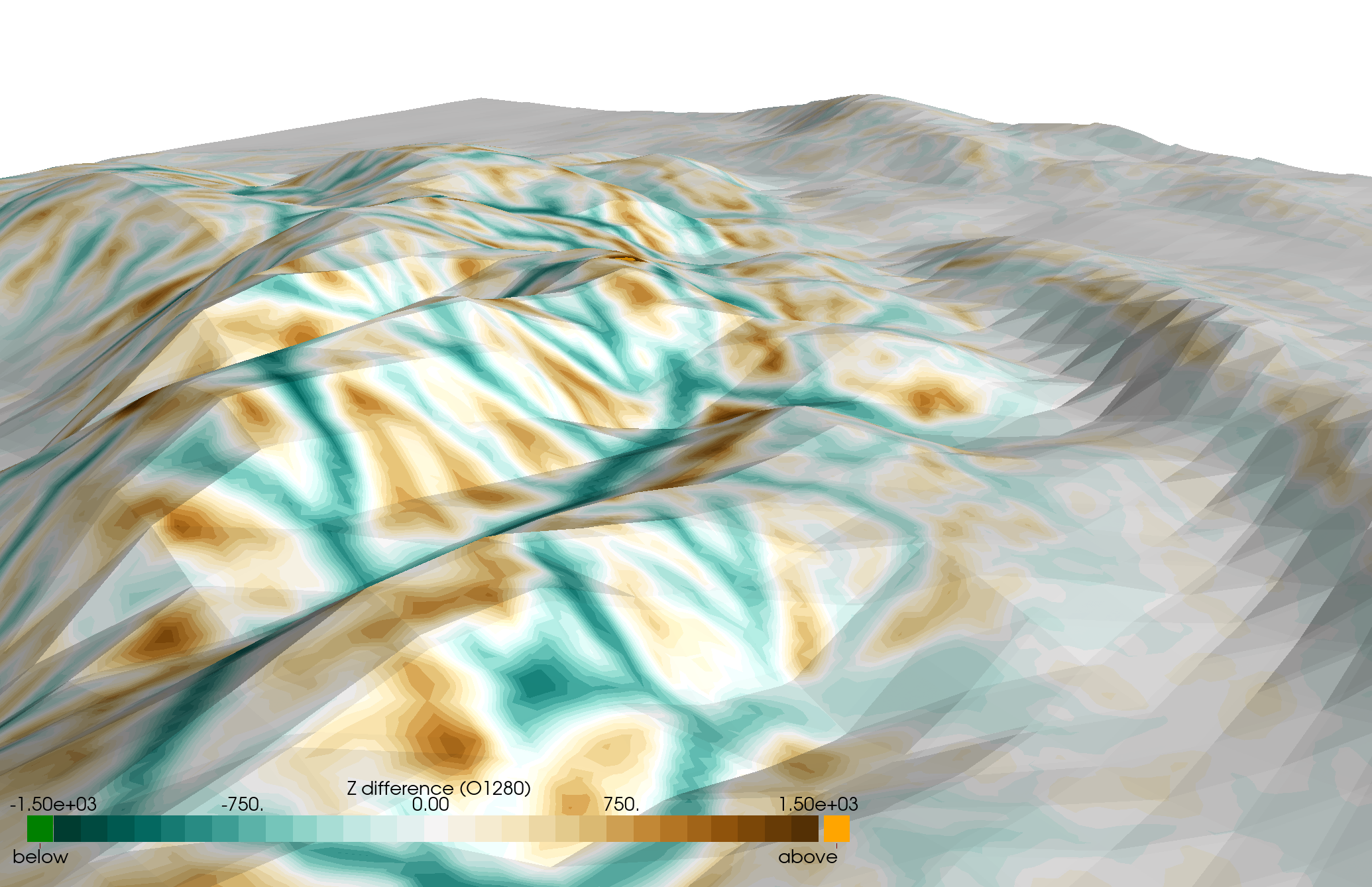}
        };
        \node[circle, inner sep=0pt, anchor=north west, fill=white, minimum size = 5mm] at (0.1, -0.1) {(b)};
    \end{tikzpicture} 
    \caption{Visualizing two terrains simultaneously. (a) High-res orography occludes low-res geometry. (b) Color coding elevation difference on the low-res domain helps to display positive and negative offsets equally.}
    \label{fig:maps-two-terrain}
\end{figure}

To show the height differences between the low- and high-res terrain, both geometries can be visualized simultaneously, with the common perceptional problems arising from such a visualization like occlusions and clutter (see \autoref{fig:maps-two-terrain} (a)). 
Note that in our use case, the low-res terrain field is needed to show the relation between simulated temperatures and either measured temperatures at stations or differences between measured and corrected temperatures at stations. The high-res field is needed to show the relation between station temperature mismatches and high-res orographic features. Thus, the domain expert usually does not use the option to show both terrains in one single view. Overall, a direct depiction of the low-res terrain can be preferable due to its simpler structure. Then, to indicate the high-res orography, height differences can be encoded via color; this is shown in \autoref{fig:maps-two-terrain} (b).

\subsection{Atmosphere layer}

The user can visualize atmospheric variables in a volumetric layer over the terrain via the atmosphere layer. This functionality was deemed important by domain experts because it shows the relationship between lapse rate, orographic features, and the 3D temperature profiles, which are predicted by the numerical weather model. 

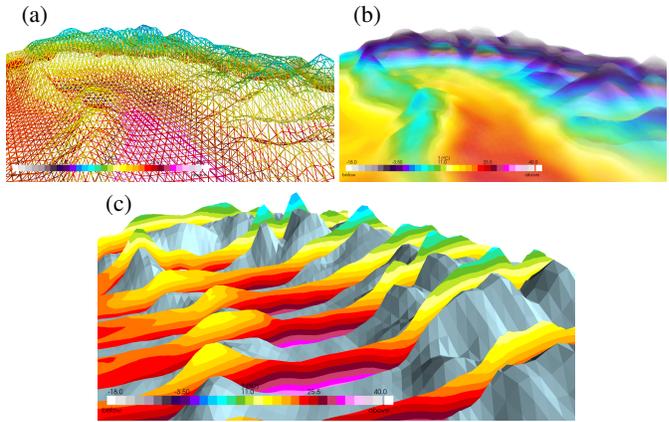
\begin{figure}[t]
    \centering
    \begin{tikzpicture}
        \node[inner sep=0pt] (highres) at (0, 0) {
        \includegraphics[width=.24\textwidth]{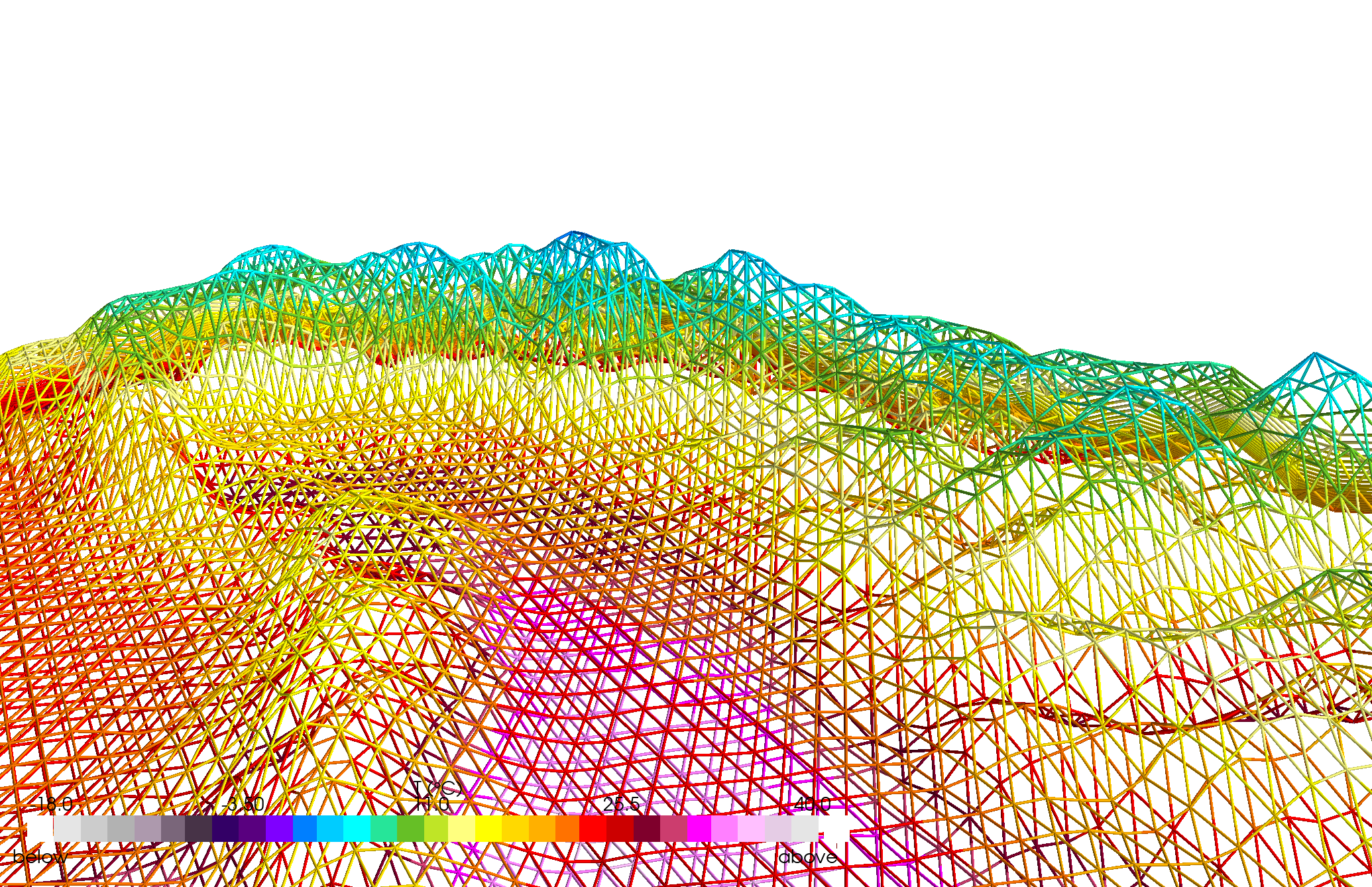}
        };
        \node at (-1.8, 0.8) {(a)};
    \end{tikzpicture}
    \begin{tikzpicture}
        \node[inner sep=0pt] (highres) at (0, 0) {
        \includegraphics[width=.24\textwidth]{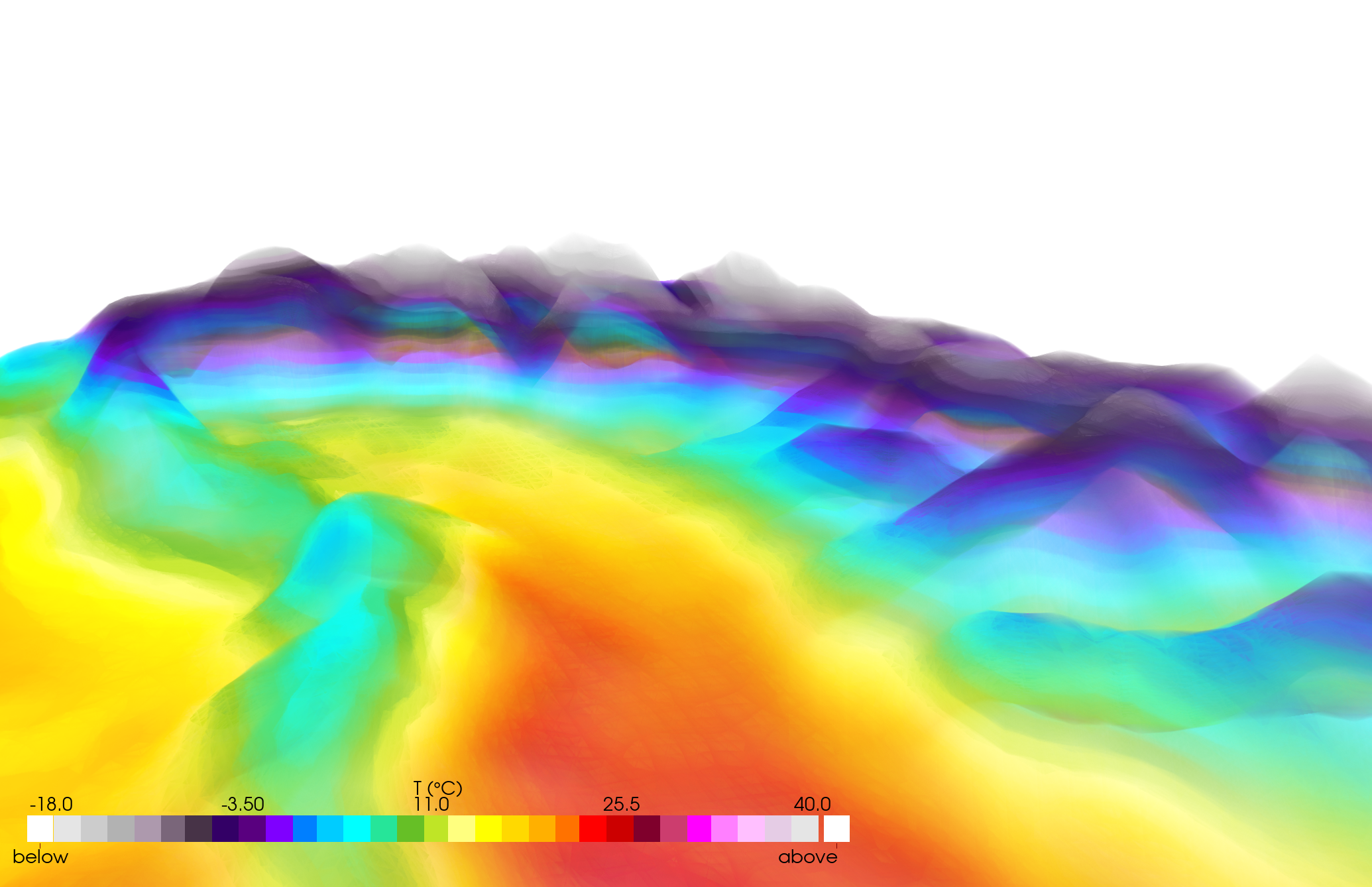}
        };
        \node at (-1.8, 0.8) {(b)};
    \end{tikzpicture}\\
    \begin{tikzpicture}
        \node[inner sep=0pt, anchor=north west] (highres) at (0, 0) {
        \includegraphics[width=.35\textwidth]{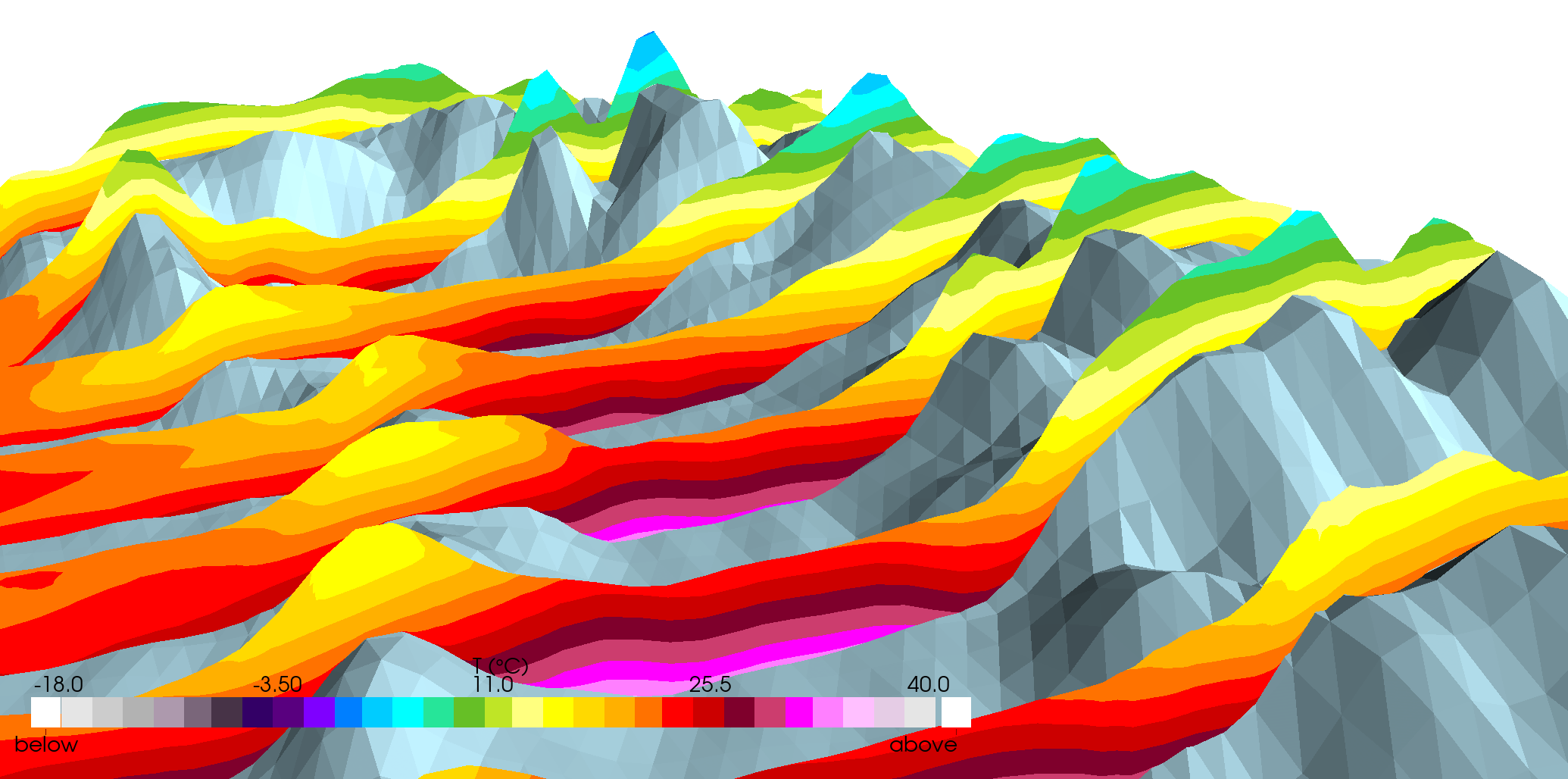}
        };
        \node[circle, inner sep=0pt, anchor=north west, fill=white, minimum size = 5mm] at (0.1, -0.1) {(c)};
    \end{tikzpicture}
        \caption{(a) The 3D terrain-following model grid as wireframe. (b) Direct volume rendering of the 3D temperature field. (c) Slicing the 3D temperature volume.}
    \label{fig:atmos-grid}
\end{figure}

To enable such visualizations, the terrain-following model grid on which a variable is given is loaded and can be shown as a colored wireframe (\autoref{fig:atmos-grid} (a)). Then, the gridded data can be rendered via volume ray-casting, with or without the terrain (\autoref{fig:atmos-grid} (b)). While volume visualization can provide a rough overview of the temperature profiles of bulk atmosphere, it hinders a fine granular analysis due to the typical attenuation and blending effects inherent to volume rendering.

To enable a more unoccluded view of the values of a gridded variable, we offer the possibility to select 2D slices oriented parallel to the 2D domain over which the terrain heightfield is given or aligned vertically along the latitudinal or longitudinal direction. Slicing is shown in \autoref{fig:atmos-grid} (c). Importantly, since the slices can be moved along their respective orthogonal direction, they can be positioned to capture certain orographic features. From the color coding of temperature on the 2D slices, the temperature distribution in the near-surface atmosphere can be revealed effectively.

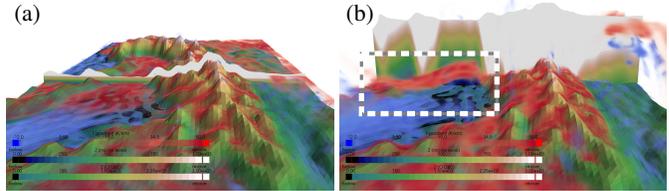
\begin{figure}[t]
    \centering
    \begin{tikzpicture}
        \node[inner sep=0pt, anchor=north west] (highres) at (0, 0) {
        \includegraphics[width=.24\textwidth]{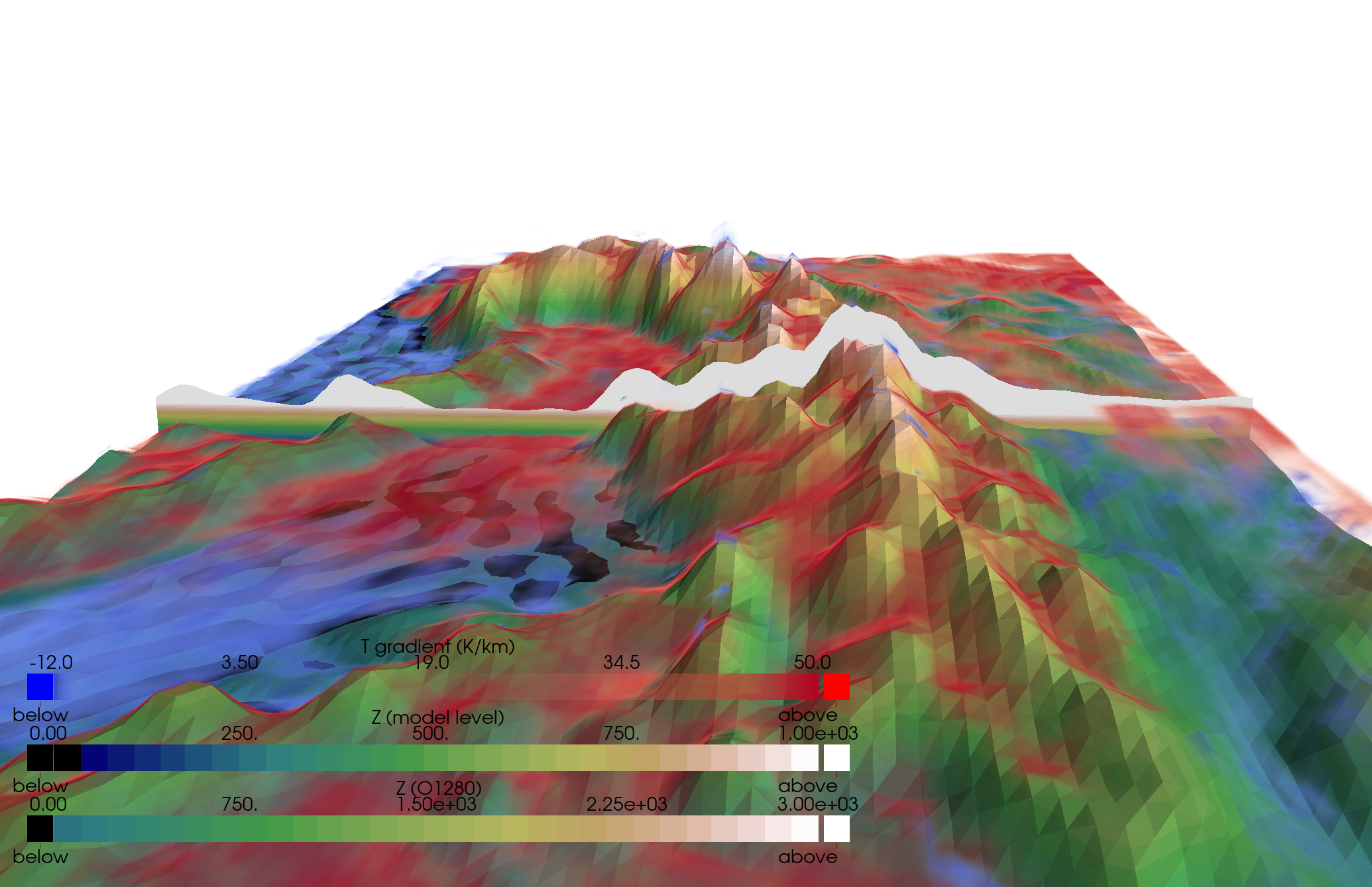}
        };
        \node[circle, inner sep=0pt, anchor=north west, fill=white, minimum size = 5mm] at (0.1, -0.3) {(a)};
    \end{tikzpicture}
    \begin{tikzpicture}
        \node[inner sep=0pt, anchor=north west] (highres) at (0, 0) {
        \includegraphics[width=.24\textwidth]{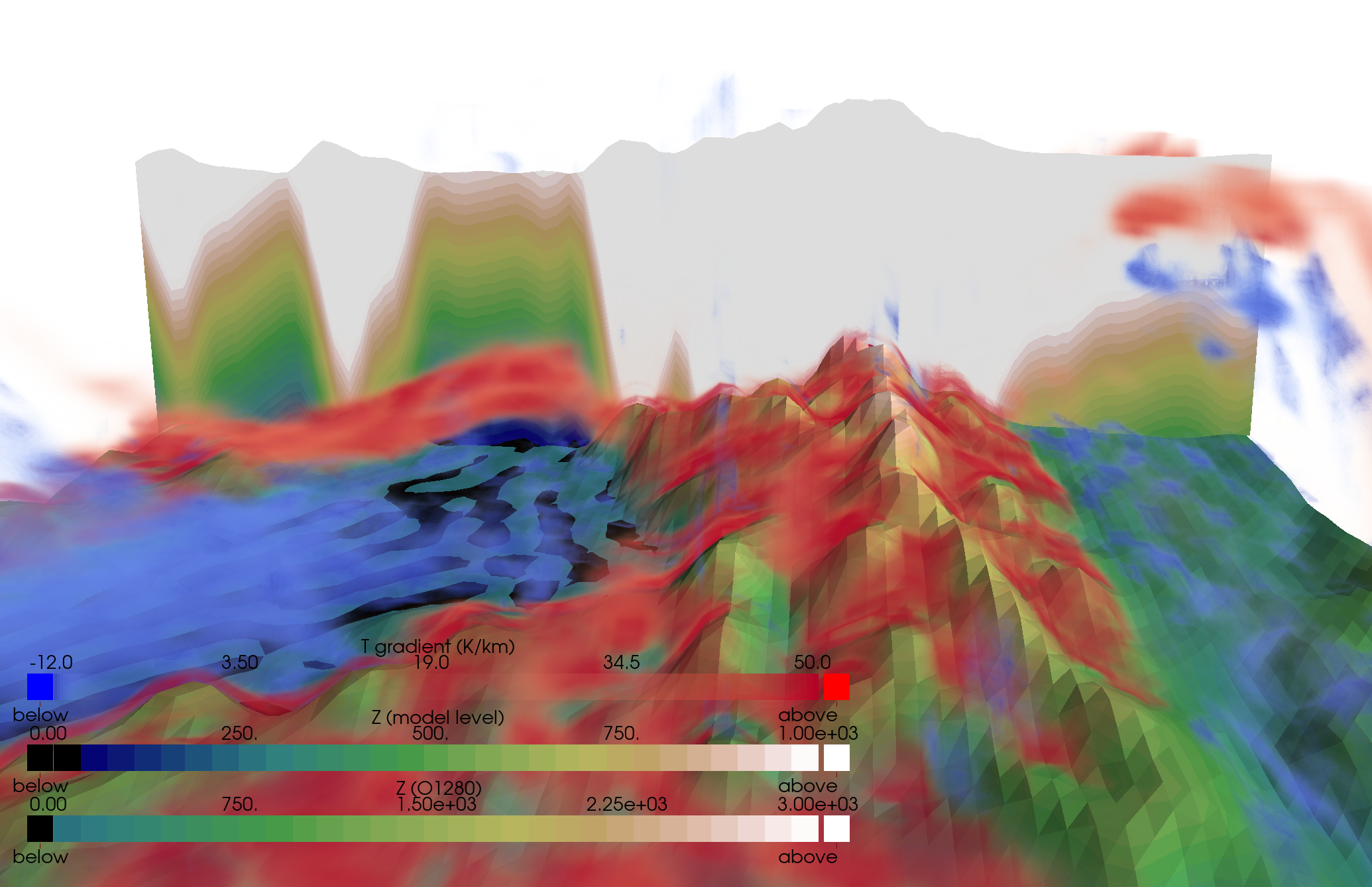}
        };
        \node[circle, inner sep=0pt, anchor=north west, fill=white, minimum size = 5mm] at (0.1, -0.3) {(b)};
        \draw[gray, very thick] (0.3,-1.) rectangle (2.1,-1.8);
        \draw[white, densely dashed, line width=2pt] (0.3,-1.) rectangle (2.1,-1.8);
    \end{tikzpicture}
        \caption{Direct volume rendering of vertical temperature gradient field, including a slice plane encoding true elevation via color. (a) No offset scaling. (b) With offset scaling. While the atmosphere layer and slice are scaled, the terrain remains unchanged. Here, the red density feature in the left-hand valley (see box) shows no connection to the valley bottom, in contrast to the red areas on top of the mountains. Colors on the reference slice indicate the red feature at around $250\,\text{m}$ altitude. 
        }
    \label{fig:atmos-slicing-grad}
\end{figure}

We provide visualizations of the vertical temperature gradient to further shed light on the local weather situation, e.g., to indicate the strength of inversions or cooling/heating effects over ground and water. \autoref{fig:atmos-slicing-grad} shows a volume rendering of this temperature gradient field overlaid on the terrain. The elevation of the sliced terrain is color-coded onto a vertical slice. Since the atmosphere layer is very narrow, the gradient distribution cannot be perceived well. To mitigate this problem, we enable offset rescaling, i.e., vertical scaling of only the extrinsic parts of the visualization, such as the volumetric grid and the slice, relative to the terrain altitude.

\subsection{Elevation summary plots}

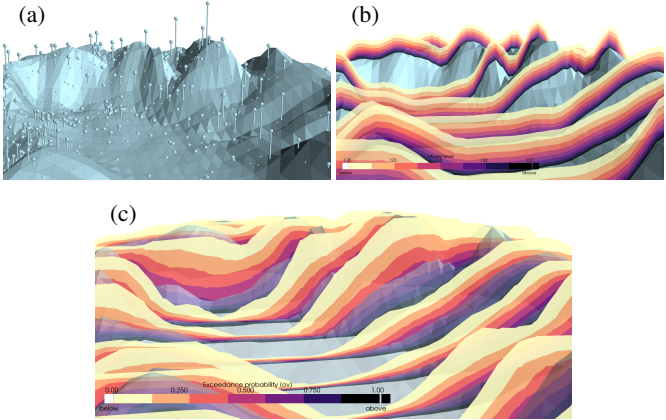
\begin{figure}
    \centering
    \begin{tikzpicture}
        \node[inner sep=0pt, anchor=north west] (highres) at (0, 0) {
            \includegraphics[width=.24\textwidth]{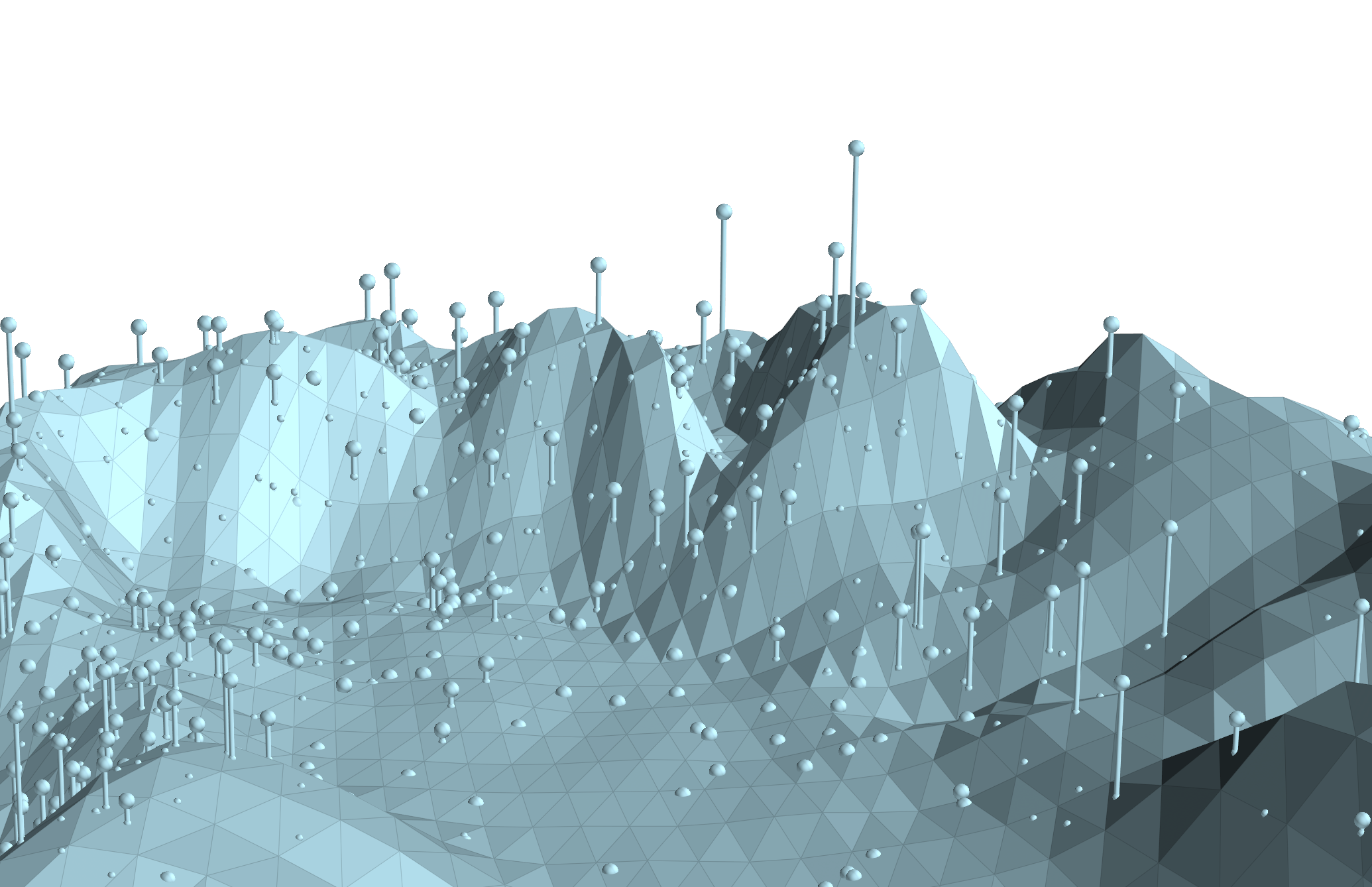}
        };
        \node[inner sep=0pt, anchor=north west] at (0.2, -0.5) {(a)};
    \end{tikzpicture}
    \begin{tikzpicture}
        \node[inner sep=0pt, anchor=north west] (highres) at (0, 0) {
            \includegraphics[width=.24\textwidth]{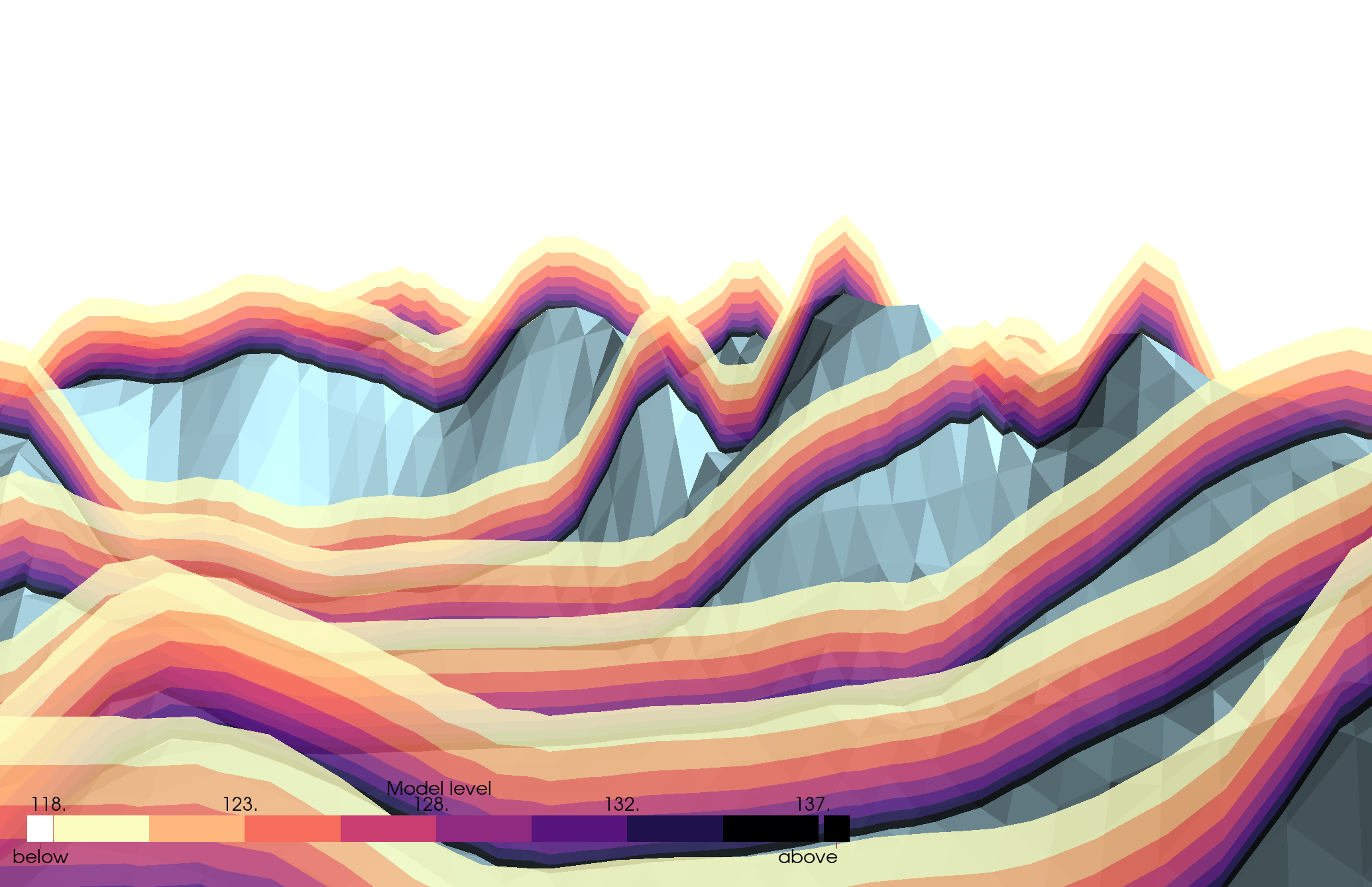}
        };
        \node[inner sep=0pt, anchor=north west] at (0.2, -0.5) {(b)};
    \end{tikzpicture}
    \begin{tikzpicture}
        \node[inner sep=0pt, anchor=north west] (highres) at (0, 0) {
                \includegraphics[width=.35\textwidth]{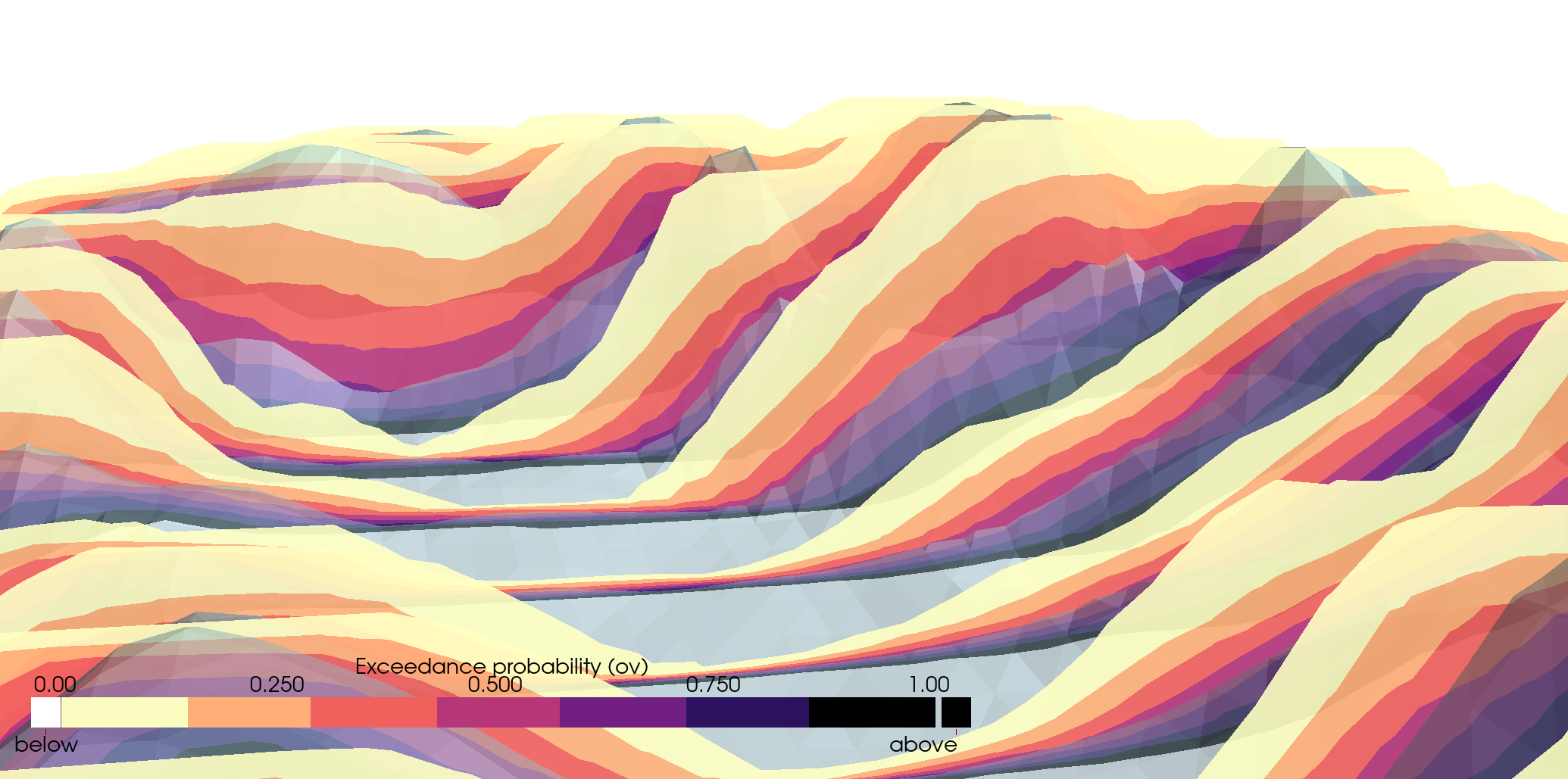}
        };
        \node[inner sep=0pt, anchor=north west] at (0.2, -0.3) {(c)};
    \end{tikzpicture}
    \caption{Overview of a vertical reference frame relative to the model orography. (a) station sites, (b) terrain-following model levels, and (c) elevation summary quantile levels with a summary radius of 60 km. The vertical coordinate of the volumetric grids is shown as color code on vertical slices through the volume. The elevation volumes stretch on both sides of the terrain, whereas the model levels are located exclusively on the upper side.}
    \label{fig:vertical-ref}
\end{figure}

Both the visualization of stations over the terrain and vertical slices through the 3D temperature field can be seen as a vertical reference indicator relative to the model orography. We provide another type of visualization in a reference frame that is defined by an elevation quantile coordinate. We call these \emph{elevation summary plots} (see \autoref{fig:vertical-ref}). 

Elevation summary plots share similarity with confidence surfaces indicating quantiles with respect to a mean surface. Such approaches typically build upon the presence of a stochastic uncertainty model or a set of ensembles from which the required statistics can be computed. In our scenario, we derive such plots for either the low- or the high-res terrain to assess respectively the elevation statistics of the lapse rate algorithm or to examine the expected sub-grid variability below the resolution of the low-res terrain.

Firstly, the user sets a search radius in units of kilometer to be considered in the statistics computation, as well as the number of quantiles to compute. For each vertex of the low-res terrain, all vertices from the so-called summary grid (the high-res or the low-res terrain surface) that are closer than the specified radius are retrieved. Note here that the distance between vertices is considered in the 2D domain over which the terrain is defined. From the elevation values of all retrieved vertices, the minimum and maximum elevation, as well as uniformly spaced quantiles of the elevation distribution according to the user-set value are computed. The quantiles are interpreted as z-coordinates of a new volumetric grid (different from the model levels), on which scalar quantities can be computed and displayed using volume visualization In \autoref{fig:vertical-ref} (c), for example, the vertical extend of the volume slices is determined by the range of minimum and maximum surface elevation in a radius of 60 km around the reference point.

Quantile computation is repeated for the area covering the interquartile range from 25\% to 75\%, yielding another volumetric mesh. On this mesh, special quantiles can be shown independently as surface meshes, e.g., the median surface and iso-layers of the 25\% and 75\% quantiles (IQR bounds). By analogy with statistical box-and-whisker plots, two additional surfaces are defined through the local elevation values: 
\begin{equation}
\begin{split}
z_\text{upper} & = z(75\%) + f \cdot \text{IQR}, \\
z_\text{lower} & = z(25\%) - f \cdot \text{IQR}, \\
\text{IQR} & = z(75\%) - z(25\%)
\end{split}
\end{equation}

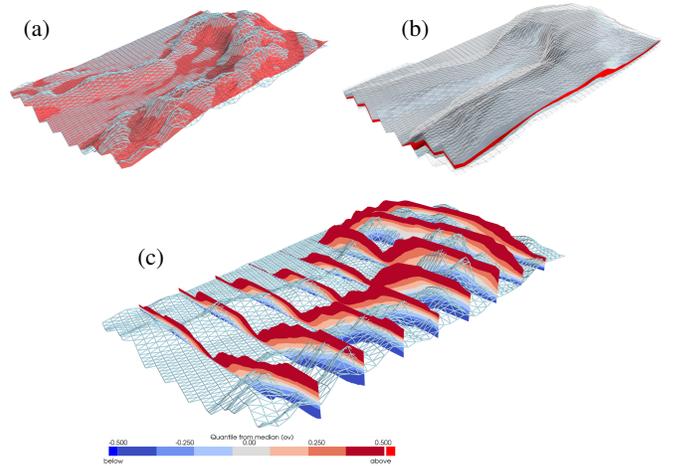
\begin{figure}
\centering
    \begin{tikzpicture}
        \node[inner sep=0pt, anchor=north west] (highres) at (0, 0) {
            \includegraphics[width=.24\textwidth]{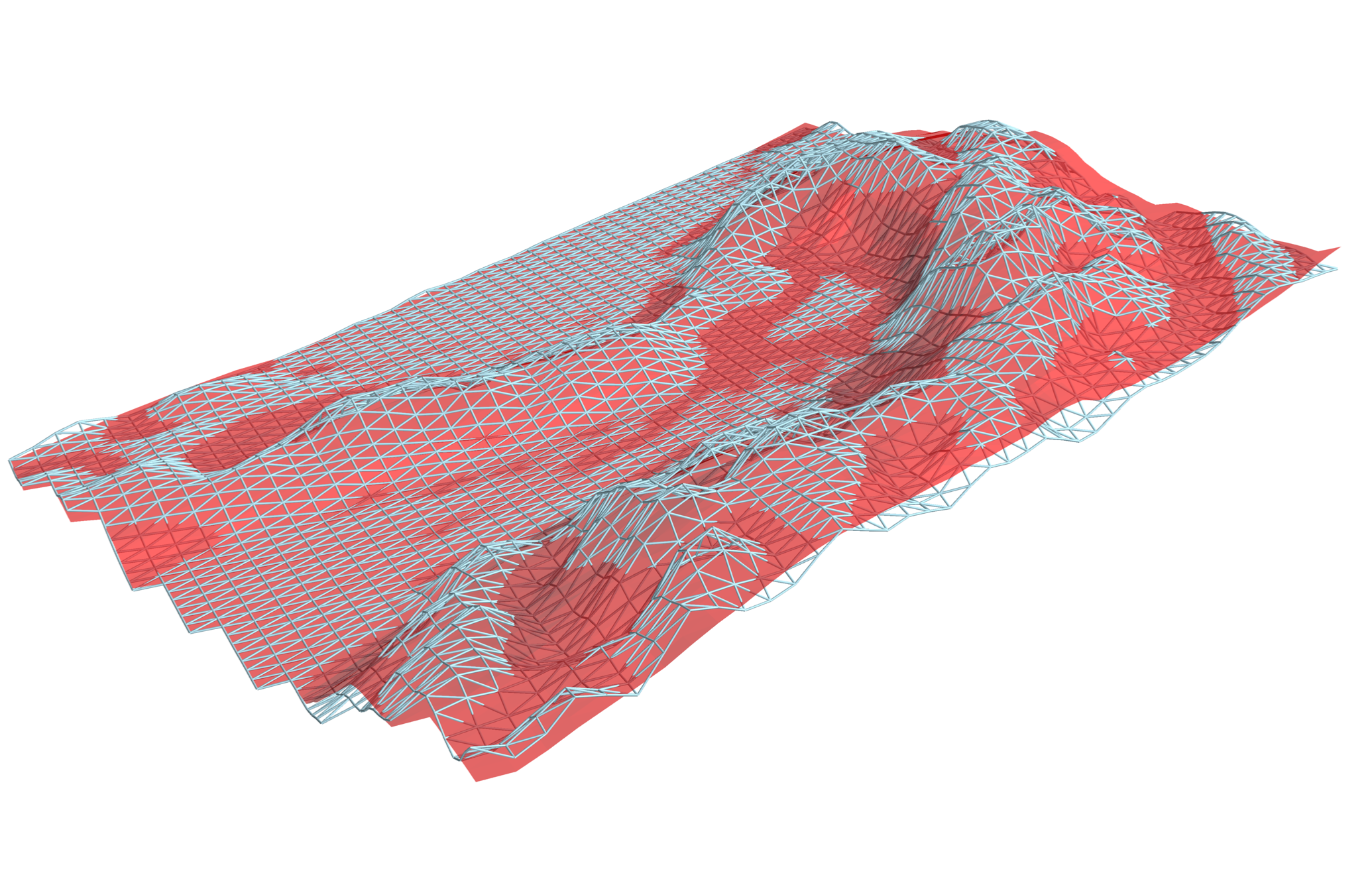}
        };
        \node[inner sep=0pt, anchor=north west] at (0.2, -0.5) {(a)};
    \end{tikzpicture}
    \begin{tikzpicture}
        \node[inner sep=0pt, anchor=north west] (highres) at (0, 0) {
            \includegraphics[width=.24\textwidth]{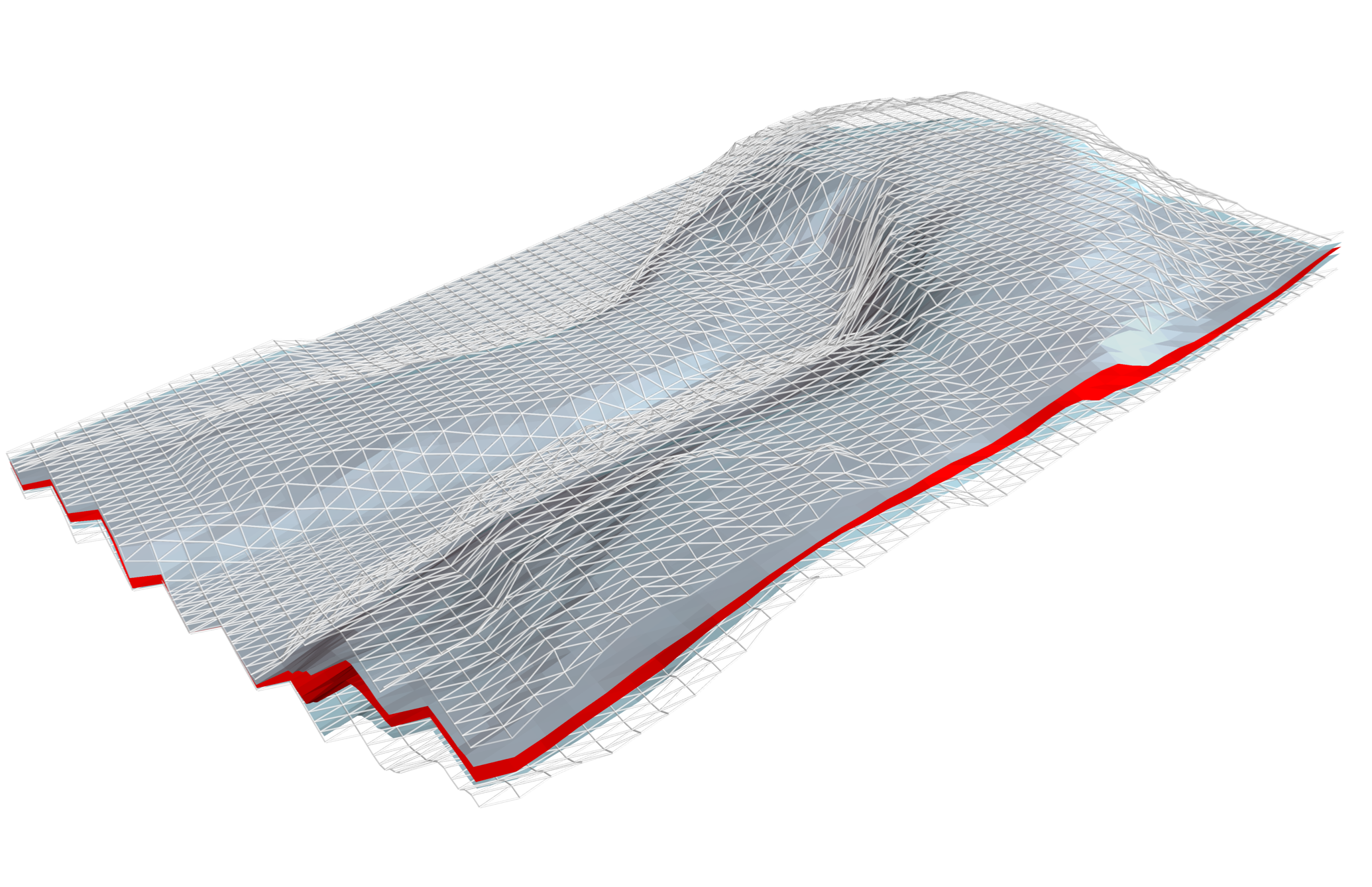}
        };
        \node[inner sep=0pt, anchor=north west] at (0.8, -0.5) {(b)};
    \end{tikzpicture}
    \begin{tikzpicture}
        \node[inner sep=0pt, anchor=north west] (highres) at (0, 0) {
            \includegraphics[width=.35\textwidth]{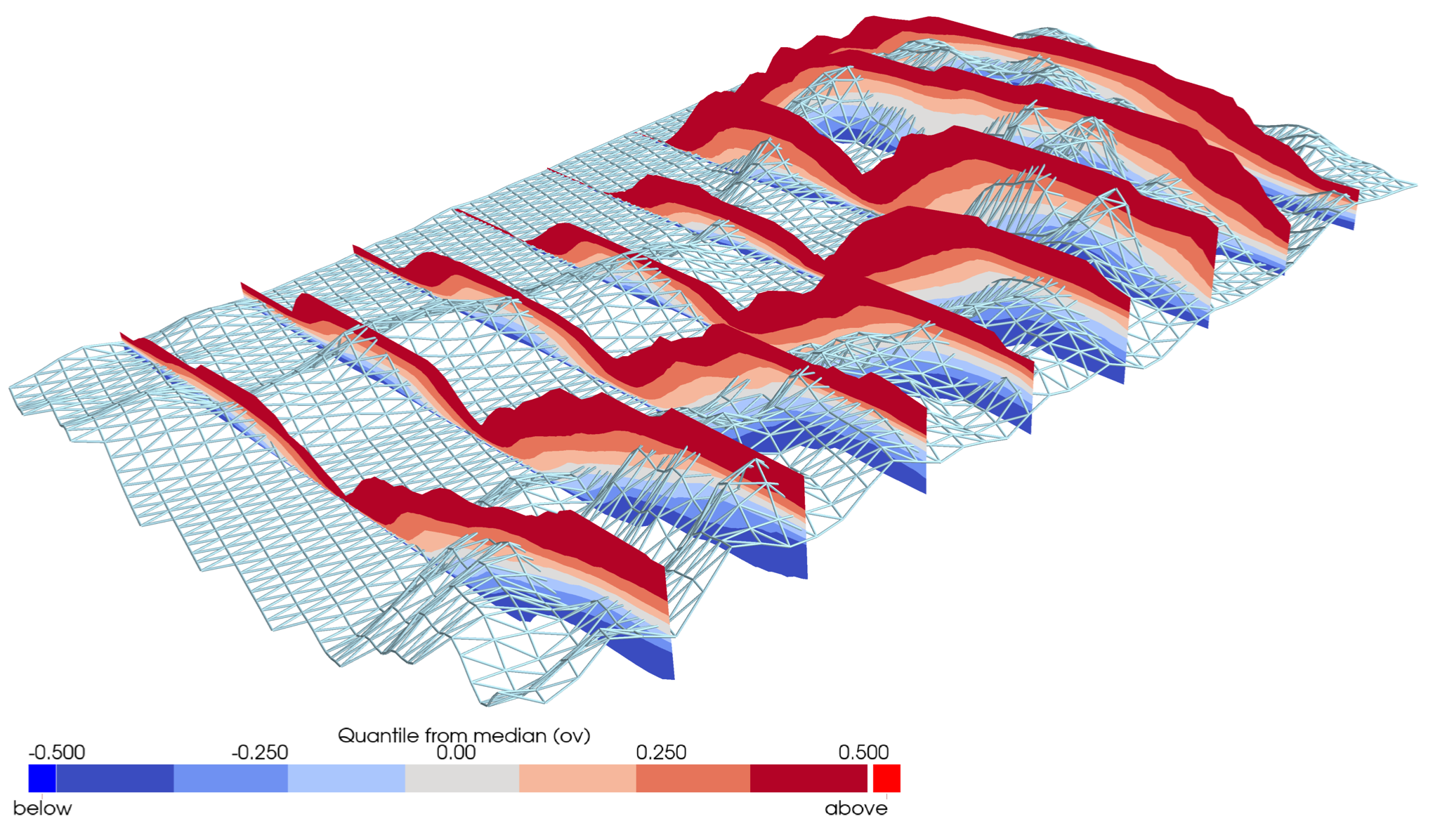}
        };
        \node[inner sep=0pt, anchor=north west] at (0.5, -0.7) {(c)};
    \end{tikzpicture}
    \caption{Elevation summary plots for the O1280 model grid with a summary radius of 60 km. (a) model grid (wireframe) with the median surface, (b) elevation boxplot with the median surface (red), IQR bounds (blue), and whiskers, and (c) model grid (wireframe) with slices through the elevation summary volume. Color on the slices indicates quantile isolevels.}
    \label{fig:elevation-summary}
\end{figure}

The upper and lower whisker heights are then given as the largest, respectively, smallest elevation sample that falls inside the range between $z_\text{upper}$ and $z_\text{lower}$. \autoref{fig:elevation-summary} illustrates the components of the elevation summary. In \autoref{fig:elevation-summary} (a), the median field is shown together with the summarized grid. In (b), the elevation summary is augmented by surfaces that visualize the IQR as well as the whisker levels. In (c), we show slices through the volume that is spanned by the minimum and maximum elevation levels of the local environments and use the slices to display statistical information.
All summary elements can be rendered jointly with stations or high-res orography, to obtain a notion of outliers in terms of local orography. For instance, the surface summary can be rendered jointly with the prediction error encoded on station sites to find stations that have high errors due to deviation.  

The provided visualization methods can be used to assess the quality of the current lapse rate scheme (see \autoref{sec:visana}) by showing the error between the corrected temperatures at stations and the temperature that was measured at these sites. As we will show in our case study, the constant lapse rate scheme introduces significant deviations to the measured temperatures. Thus, in the following we suggest an alternative lapse rate scheme that results in improved temperature corrections. 

\subsection{Improved lapse rate scheme}

Instead of a fixed global lapse rate, we propose to compute a local lapse rate from the current model data. 
The method starts with estimating the local lapse rate at the vertices (gridpoints) of the low-res model grid. To do this, the domain expert sets a radius that corresponds to a "reasonable" scale in terms of local weather conditions, typically a value between 40 and 60 kilometers. From all vertices (excluding non-land vertices) within the region indicated by the set radius, all 2 m temperatures and elevation values are collected. Through a linear model that fits predicted temperature as a function of elevation, we obtain a local weight coefficient relating elevation to temperature. This coefficient is used as the local lapse rate estimate. To avoid spatial discontinuities as a consequence of using a hard cutoff radius, a Gaussian weighting scheme is applied, which assigns a higher weight to the samples close to the reference location. Validity of the local linear model is assessed via the coefficient of determination, also called $R^2$ score. 
To guarantee proper convergence of the estimator, lapse rates are estimated only for grid vertices, which have at least 20 non-sea grid vertices within their radial neighborhood. For other sites, the scheme reverts to the default lapse rate.

\begin{figure*}
    \centering
    \begin{tikzpicture}
        \node[inner sep=0pt, anchor=north west] (highres) at (0, 0) {
            \includegraphics[width=.33\textwidth]{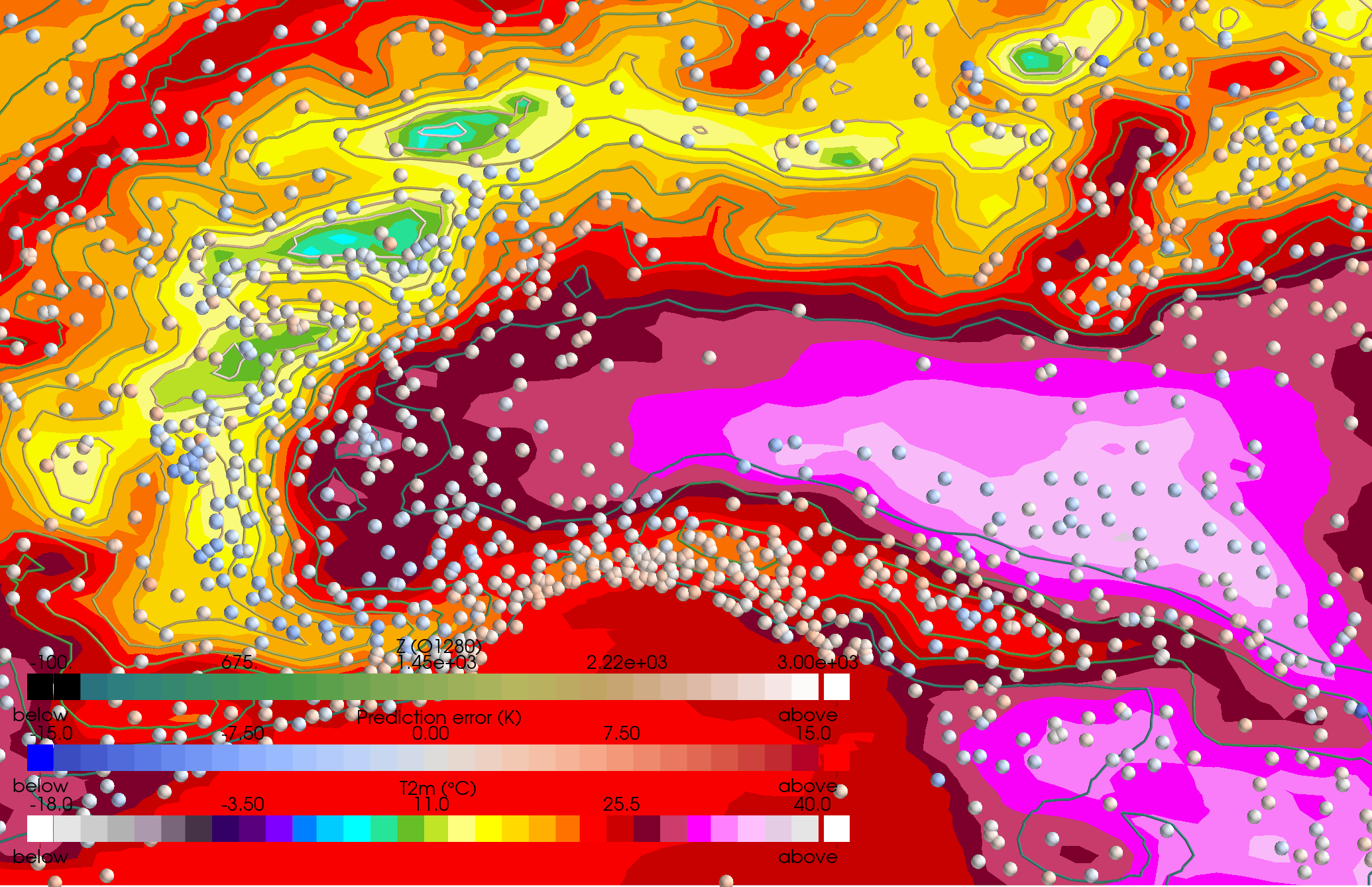}
        };
        \node[circle, inner sep=0pt, anchor=north west, fill=white, minimum size = 5mm] at (0.1, -0.1) {(a)};
    \end{tikzpicture}
    \begin{tikzpicture}
        \node[inner sep=0pt, anchor=north west] (highres) at (0, 0) {
            \includegraphics[width=.33\textwidth]{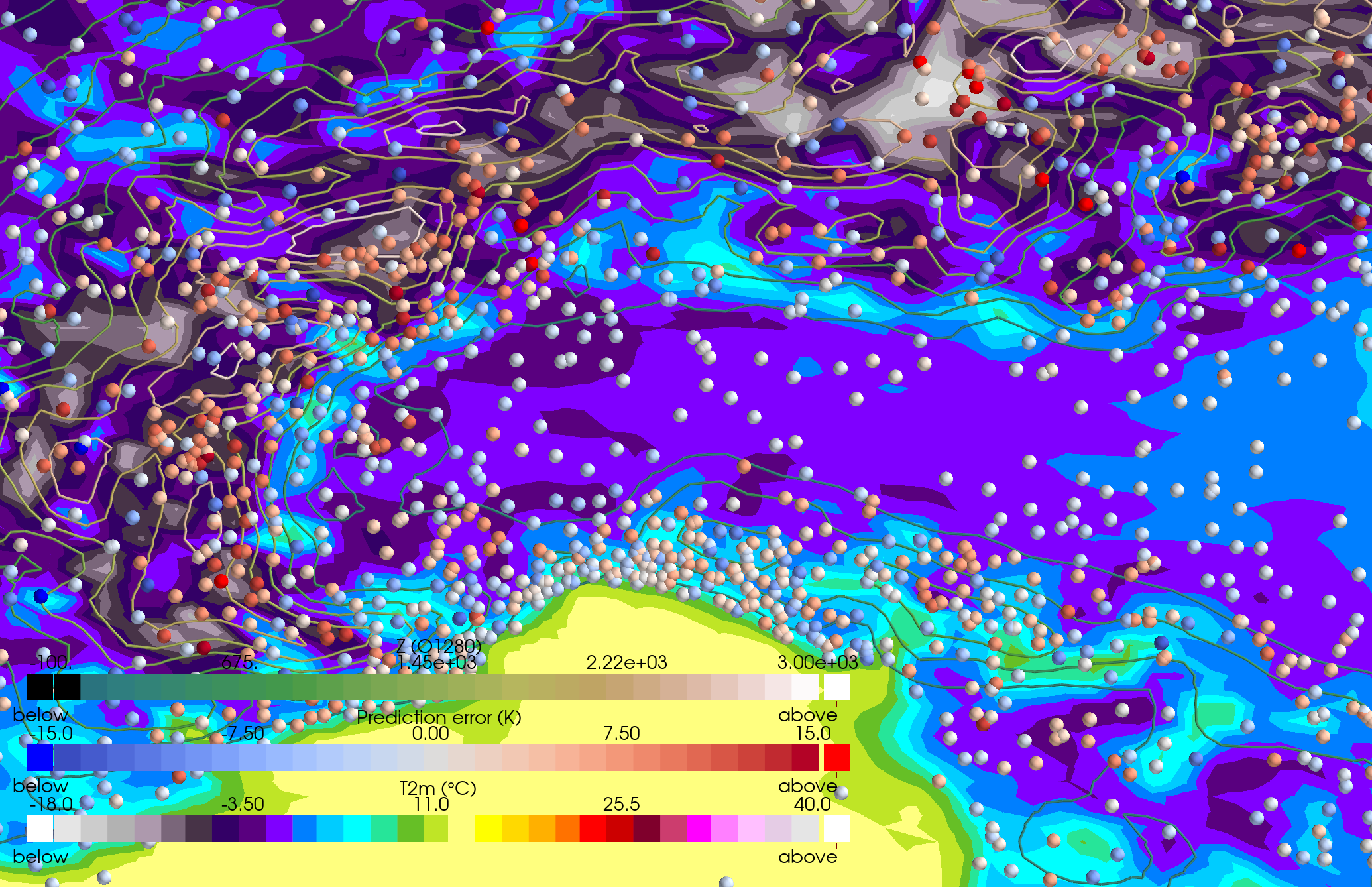}
        };
        \node[circle, inner sep=0pt, anchor=north west, fill=white, minimum size = 5mm] at (0.1, -0.1) {(b)};
    \end{tikzpicture}
    \begin{tikzpicture}
        \node[inner sep=0pt, anchor=north west] (highres) at (0, 0) {
            \includegraphics[width=.33\textwidth]{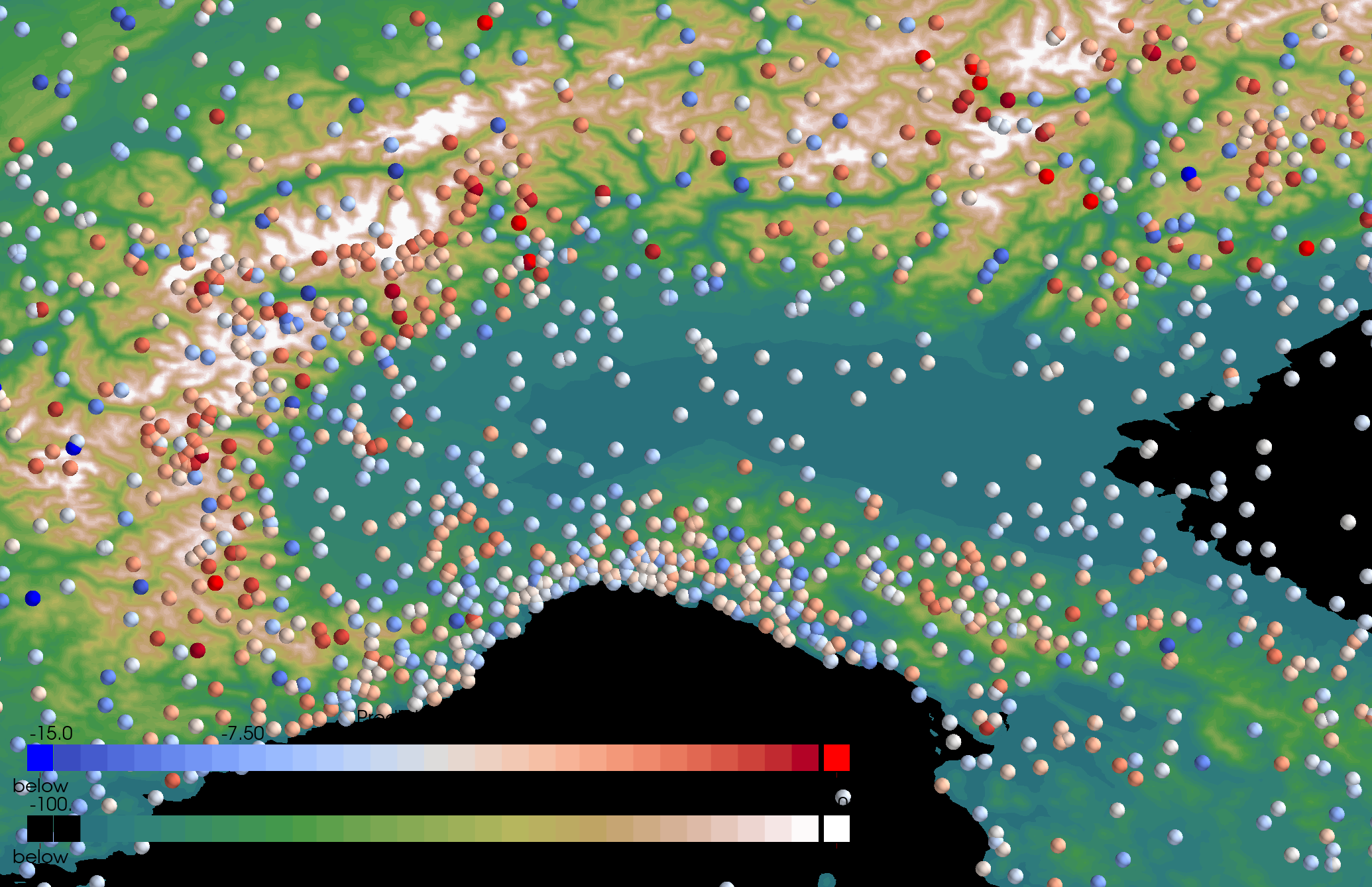}
        };
        \node[circle, inner sep=0pt, anchor=north west, fill=white, minimum size = 5mm] at (0.1, -0.1) {(c)};
        \draw[white, densely dashed, line width=2pt] (0.4,-0.6) rectangle (2,-2.5);
    \end{tikzpicture}
    \caption{Overview of prediction errors and model surface temperatures for (a) summer and (b) winter case study, and (c) shows the winter prediction errors in the context of the model orography. The box indicates the focus region of figures \ref{fig:winter-detail-view} and \ref{fig:case-temp-slices}.}
    \label{fig:case-overview}
\end{figure*}
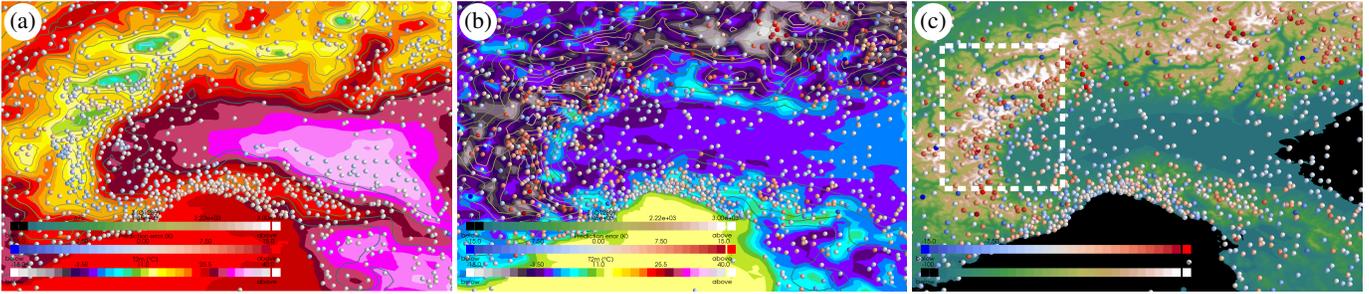

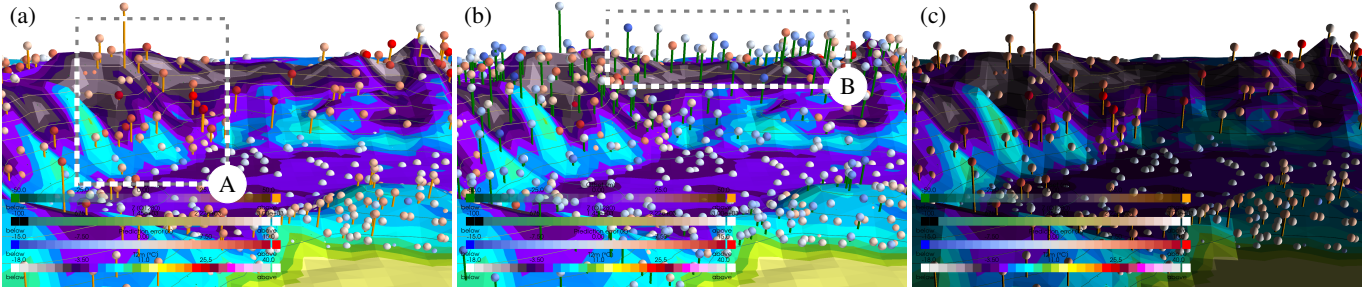
\begin{figure*}
    \centering
    \begin{tikzpicture}
        \node[inner sep=0pt, anchor=north west] (highres) at (0, 0) {
            \includegraphics[width=.33\textwidth]{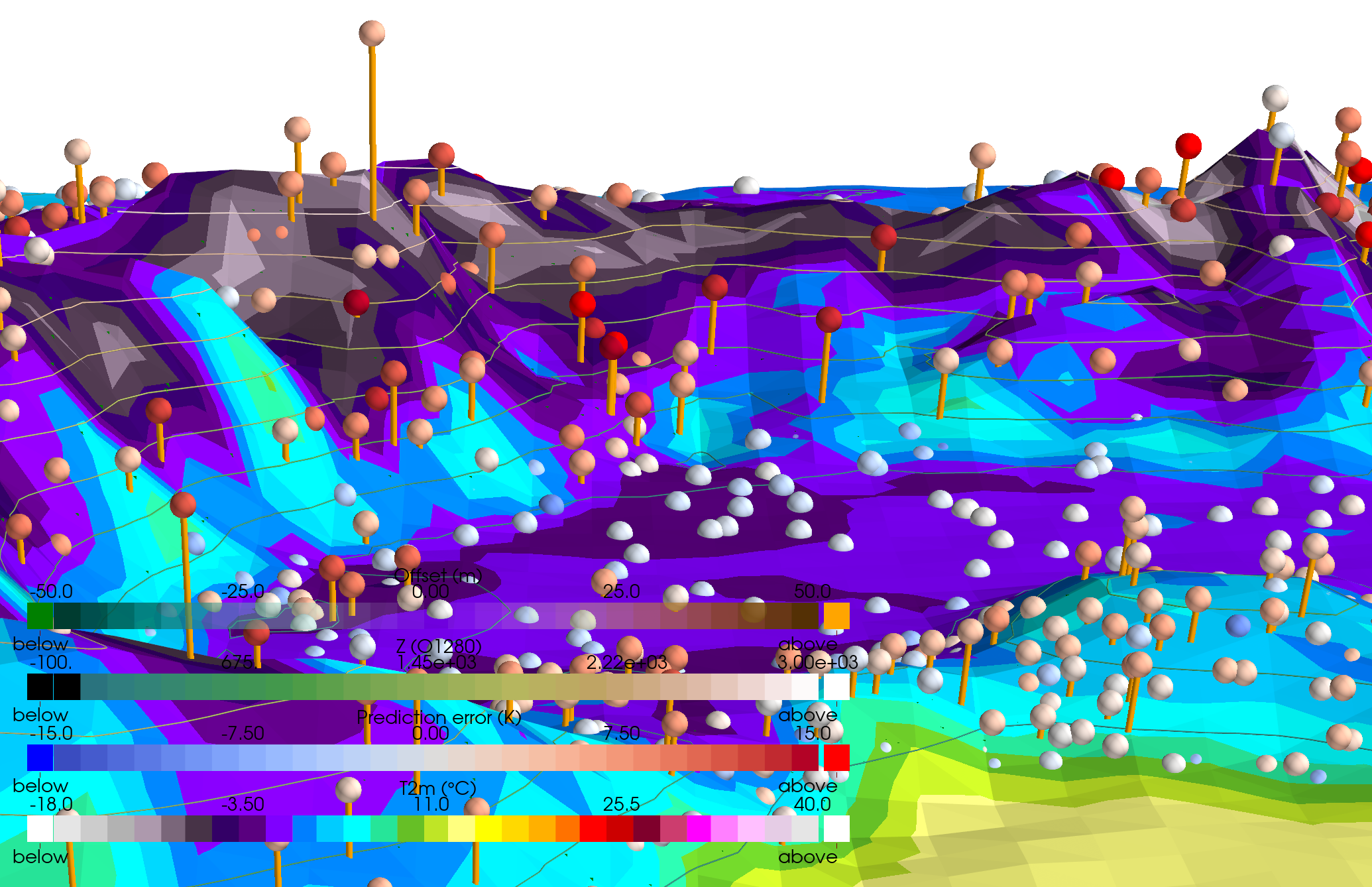}
        };
        \node[circle, inner sep=0pt, anchor=north west, fill=white, minimum size = 5mm] at (0.1, -0.1) {(a)};
        \draw[gray, very thick] (1.,-0.3) rectangle (3,-2.5);
        \draw[white, densely dashed, line width=2pt] (1.,-0.3) rectangle (3,-2.5);
        \node[circle, inner sep=0pt, fill=white, draw=white, minimum size = 5mm] at (3,-2.5) {A};
    \end{tikzpicture}
    \begin{tikzpicture}
        \node[inner sep=0pt, anchor=north west] (highres) at (0, 0) {
            \includegraphics[width=.33\textwidth]    {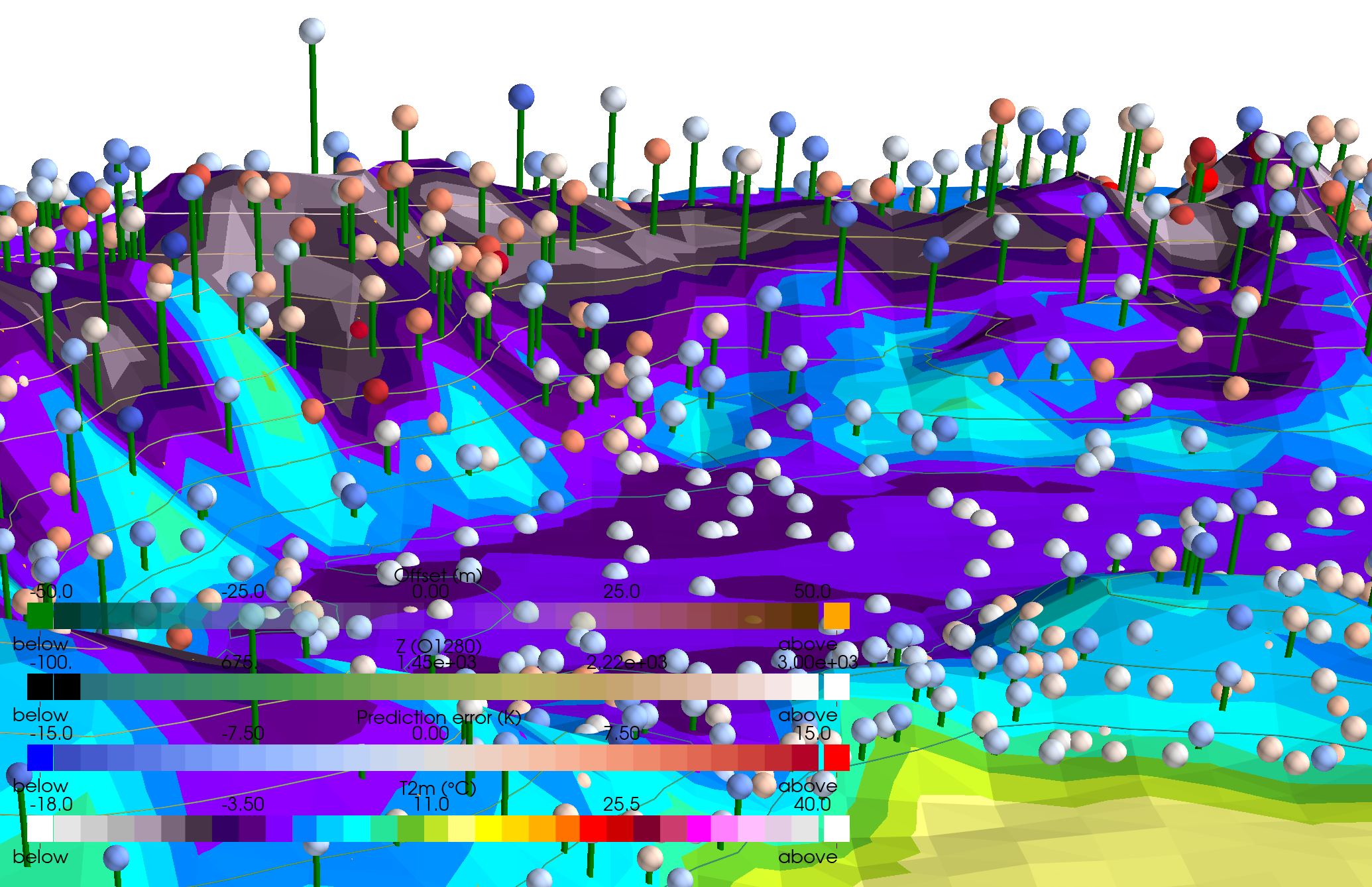}
        };
        \node[circle, inner sep=0pt, anchor=north west, fill=white, minimum size = 5mm] at (0.1, -0.1) {(b)};
        \draw[gray, very thick] (2.,-0.2) rectangle (5.2,-1.2);
        \draw[white, densely dashed, line width=2pt] (2.,-0.2) rectangle (5.2,-1.2);
        \node[circle, inner sep=0pt, draw=white, fill=white, minimum size = 5mm] at (5.2,-1.2) {B};
    \end{tikzpicture}    
    \begin{tikzpicture}
        \node[inner sep=0pt, anchor=north west] (highres) at (0, 0) {
            \includegraphics[width=.33\textwidth]{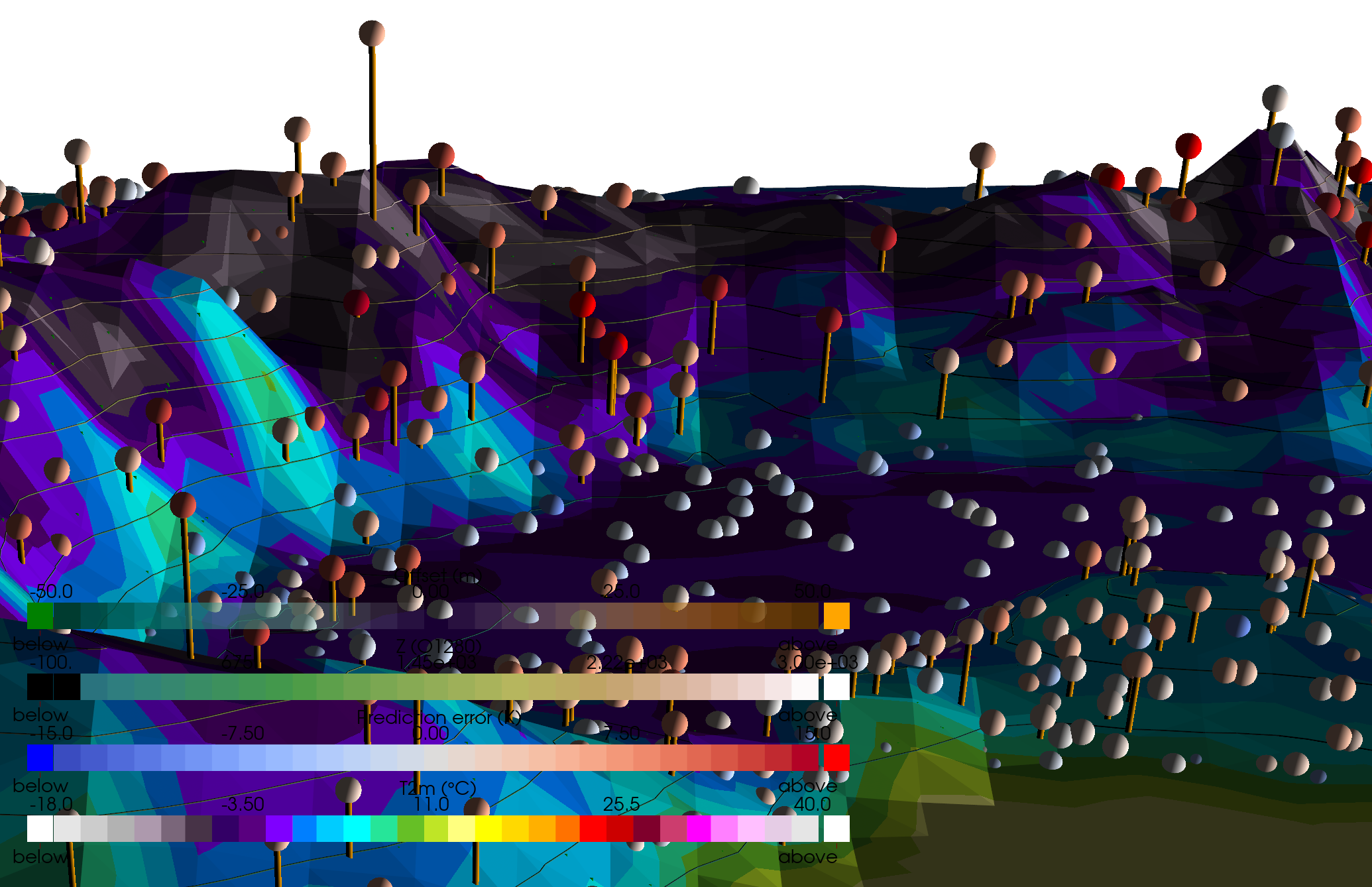}
        };
        \node[circle, inner sep=0pt, anchor=north west, fill=white, minimum size = 5mm] at (0.1, -0.1) {(c)};
    \end{tikzpicture}    
    \caption{Northward overview of prediction errors and model surface temperature for the winter case, (a) in regular view on the mountain stations, (b) with inverse station offset to display valley stations, and (c) with solar lighting, highlighting vertical structures in the model temperature field.}
    \label{fig:winter-detail-view}
\end{figure*}

To circumvent the potentially harmful effects of extreme lapse rate estimates, min-max clamping is applied, in which the upper and lower bounds depend on $R^2$ of the local linear model. We provide an interface to parametrize upper and lower bounds $b_\text{up/low}$ as a ramp function:
\begin{equation}
    b_{\cdot} = \left\{
    \begin{array}{ll}
         c_\text{lower} & \text{if } R^2 < r_\text{lower}\text{,} \\ 
         c_\text{upper} & \text{if } R^2 \geq r_\text{upper}\text{,} \\
         c_\text{lower} + \frac{c_\text{upper} - c_\text{lower}}{r_\text{upper} - r_\text{lower}} (R^2 - r_\text{lower}) & \text{else.}
    \end{array}
    \right.
    \label{eq:clamps}
\end{equation}

Default values for the parameters are selected empirically (see \autoref{sec:quality}) and are set to $(c_\text{lower}, c_\text{upper}, r_\text{lower}, r_\text{upper}) = (20 \text{ K/km}, 50 \text{ K/km}, 0, 1)$ for the upper lapse rate bound, and $(-6.5 \text{ K/km}, -11 \text{ K/km}, 0.75, 0.95)$ for the lower bound. 
Based on the estimated lapse rate, a corrected prediction for each location of interest (grid vertex or station site) is then obtained  as $T_\text{site} = T_\text{model} + \gamma (z_\text{site} -z_\text{model})$, wherein $T_\text{model}$, $z_\text{model}$,  and $\gamma$ are 2 m temperature model prediction, model grid elevation, the local lapse rate estimate at the closest vertex in the model grid, and $z_\text{site}$ is the station altitude.

\section{Use case}

\subsection{Visual analysis of near-surface temperature correction \label{sec:visana}}

To obtain an impression of the limitations of the constant lapse rate approach (\textbf{Q1 + Q2}), two timestamps are selected, for which model predictions are visualized jointly with station observations. The cases are indicative, over central Europe, of the two ends of the lapse rate spectrum. That is a meteorologically less stable situation, from a summer afternoon, July 12, 2021, 1500 UTC, and a meteorologically more stable situation from an anticyclonic winter morning, December 19, 2021, 0600 UTC. Both featured light near-surface winds, as stronger winds tend to promote more standard lapse rates via turbulent mixing. Furthermore, due to the low-level inversion / high stability, the winter case had proved especially challenging for operational model 2 m temperature forecasts and reanalysis (Figure 3 in \cite{highlander}).

\autoref{fig:case-overview} (a) and (b) display top-down map views for both cases. A constant-altitude projection is used to enable the display of stations above and below the terrain at the same time. The color of station indicators encodes the prediction errors of the standard lapse rate scheme at the station site. Orography is displayed as isocontours to reveal relations between temperature distribution and terrain properties. Station indicators and terrain isocontours are shaded as volumetric spheres and tubes, respectively, to improve their perception in front of the bright model temperature map in the background. Temperature isocontours are visible as color boundaries in the temperature color map. 

In the summer case, temperature isocontours adhere closely to the contour lines of the orography field, both in orientation and spacing. This indicates that the constant lapse rate assumption is a good approximation of the weather situation. Correspondingly, the prediction errors, encoded in sphere color, are generally no more than a few Kelvin.

In the winter case, the prediction errors are much larger, especially in the mountainous areas. This suggests that the constant lapse rate scheme does not yield skillful predictions here. The deviation of the temperature conditions from the idealized model is also confirmed by the relation of the temperature and orography isocontours. In parts, the isocontours intersect orthogonally, indicating that temperatures do not change with altitude at all. To examine the relation between orography and prediction errors in more detail, the model temperature map is replaced with a map representation of the high-res orography in \autoref{fig:case-overview} (c). The view reveals even more clearly that the magnitude of prediction errors correlates with the mountainous character of the terrain. Additionally, the user may recognize that stations with large prediction errors are located mainly in places with right-facing mountain slopes (i.e., on eastern slopes). An area where this is particularly apparent is highlighted with a white box in \autoref{fig:case-overview} (c) and examined in more detail in a 3D view in \autoref{fig:winter-detail-view}.

In \autoref{fig:winter-detail-view} the surface color is used to encode model 2 m temperature. Station sites are shown on top of tube lines, which are colored according to the elevation difference between station and model terrain. Stations with an elevation difference above 50 m are classified as convex (mountain) stations, whereas stations more than 50 m below the model orography are classified as concave (valley) stations. \autoref{fig:winter-detail-view} (a) displays station sites above the model orography, and suggests that convex stations exhibit large positive discrepancies versus the model predictions (see box A). The offset inversion feature is then used to switch to the view of \autoref{fig:winter-detail-view} (b), in which convex stations change positions with concave stations. In contrast to the convex stations, the view now suggests that valley stations on the backside of the mountain ridge (see box B) have a tendency to show lower values than predicted. 

Considering the timestamp of the weather situation, 0600 UTC, corresponding to 0700 CET -- the local time in the area of interest -- the temperature anomalies may be caused by the formation of cold air pools in mountain valleys for various reasons: e.g.\ radiative cooling overnight, katabatic drainage into the valleys, mountains casting shadows onto the valleys by day. The latter aspect can be investigated by switching to solar lighting mode, which simulates a light source coming from the direction of the solar irradiation. This option is seen in \autoref{fig:winter-detail-view} (c), indicating that this may indeed contribute to higher temperature measurements, as the sunlight is coming from the south-eastern direction.

\begin{figure*}
    \centering
    \begin{tikzpicture}
        \node[inner sep=0pt, anchor=north west] (highres) at (0, 0) {
            \includegraphics[width=.33\textwidth]{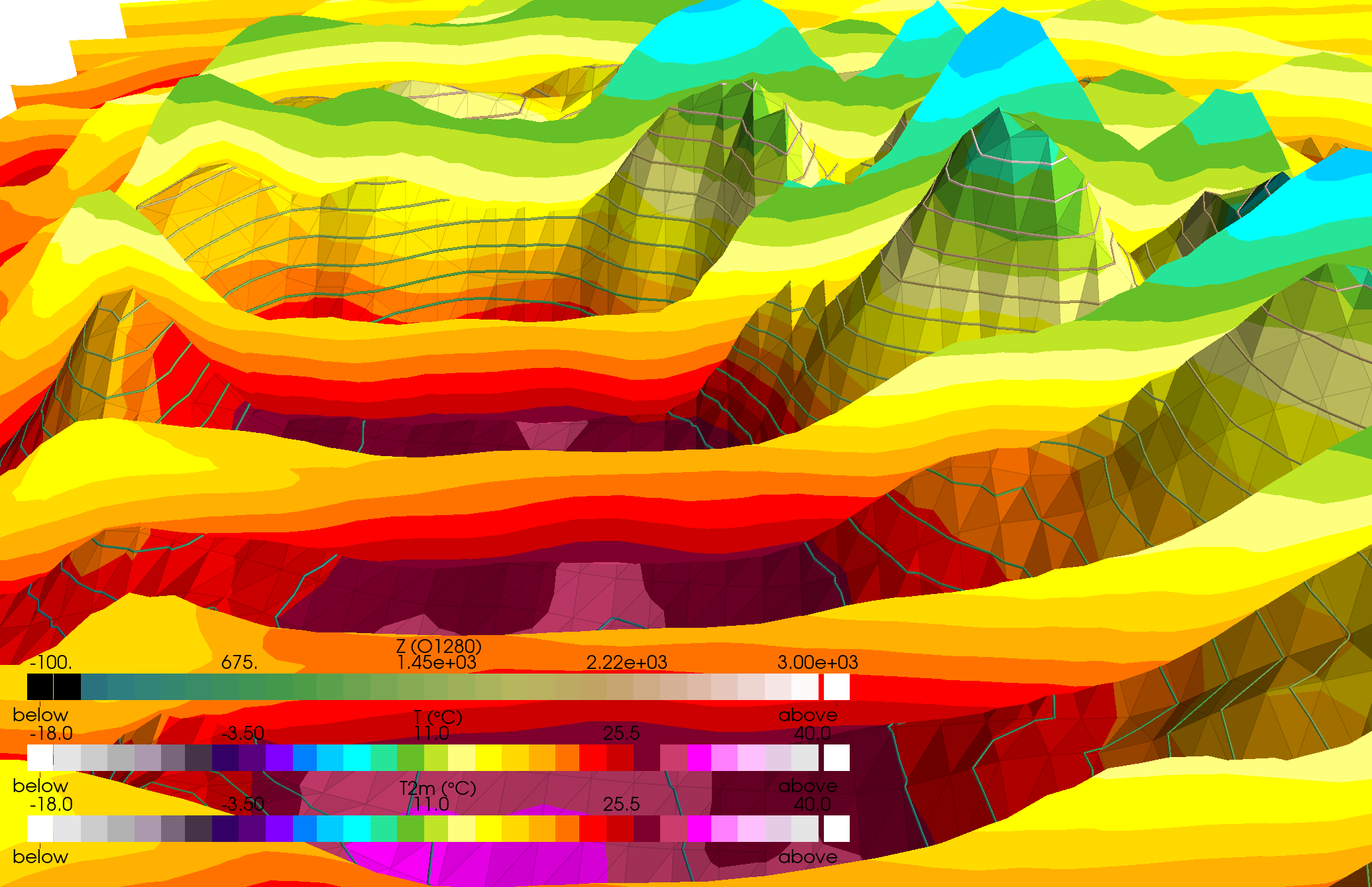}
        };
        \node[circle, inner sep=0pt, anchor=north west, fill=white, minimum size = 5mm] at (0.1, -0.1) {(a)};
        \draw[white, densely dashed, ultra thick] (3.5, -0.2) rectangle (4.5,-1.5);
                \node[circle, inner sep=0pt, fill=white, minimum size = 5mm] at (4.5,-1.5) {A1};
    \end{tikzpicture}    
    \begin{tikzpicture}
        \node[inner sep=0pt, anchor=north west] (highres) at (0, 0) {
            \includegraphics[width=.33\textwidth]{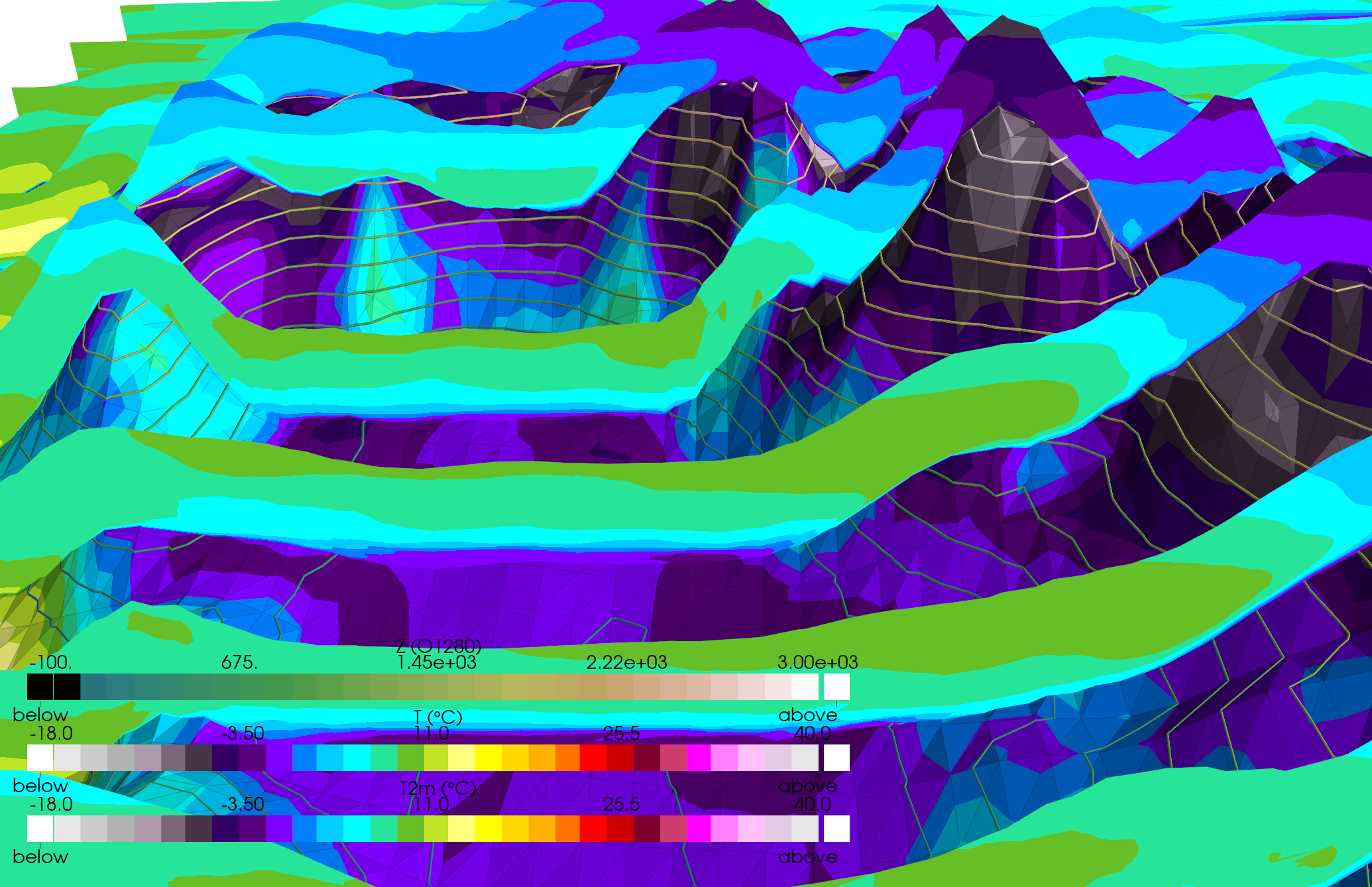}
        };
        \node[circle, inner sep=0pt, anchor=north west, fill=white, minimum size = 5mm] at (0.1, -0.1) {(b)};
        \draw[white, densely dashed, ultra thick] (3.5, -0.2) rectangle (4.5,-1.5);
        \node[circle, inner sep=0pt, fill=white, minimum size = 5mm] at (4.5,-1.5) {A2};
        \draw[white, densely dashed, ultra thick] (1.2, -1.8) rectangle (3.7,-2.6);
        \node[circle, inner sep=0pt, fill=white, minimum size = 5mm] at (3.7,-2.6) {B1};
    \end{tikzpicture}    
    \begin{tikzpicture}
        \node[inner sep=0pt, anchor=north west] (highres) at (0, 0) {
            \includegraphics[width=.33\textwidth]{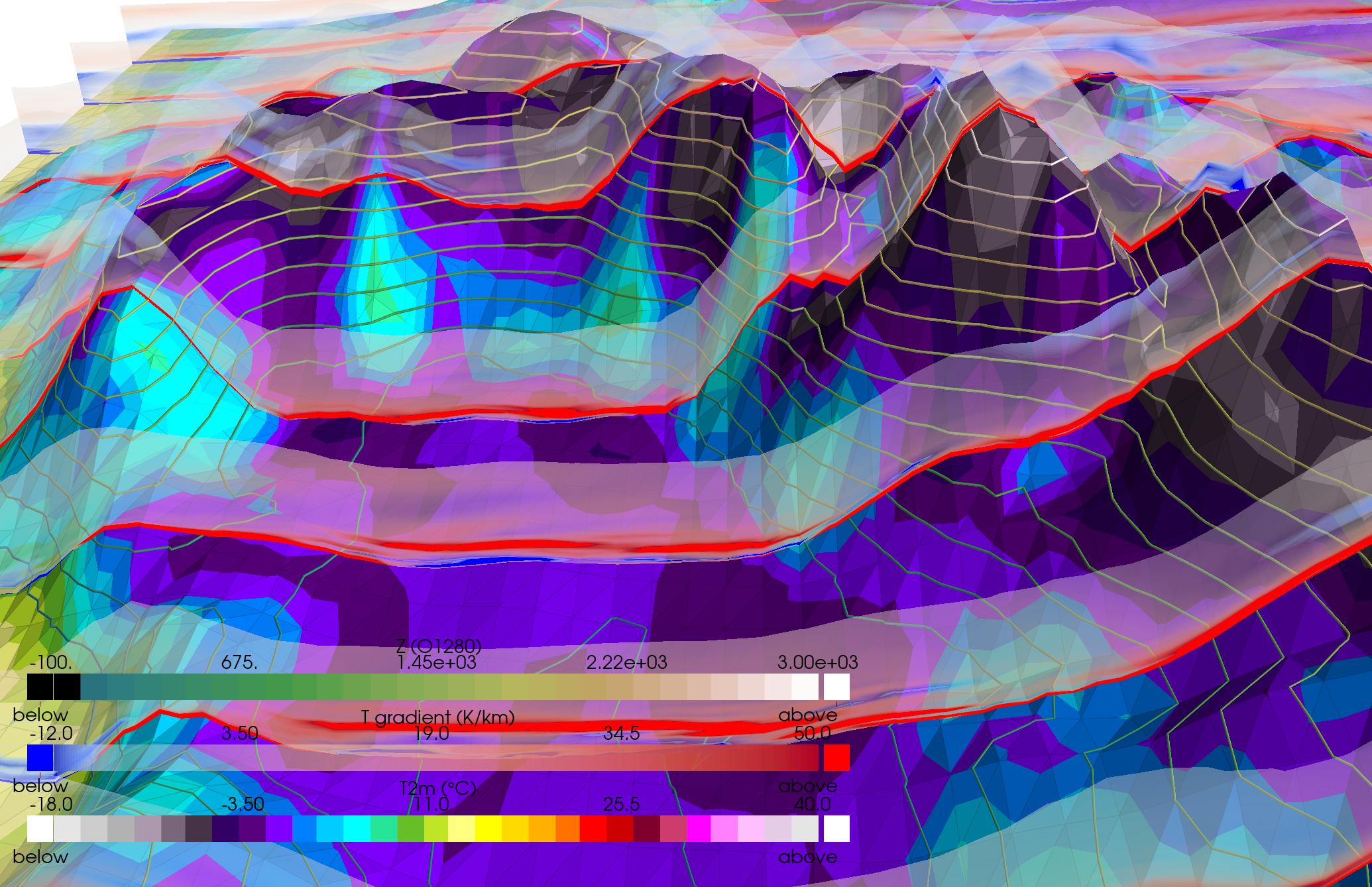}
        };
        \node[circle, inner sep=0pt, anchor=north west, fill=white, minimum size = 5mm] at (0.1, -0.1) {(c)};
        \draw[white, densely dashed, ultra thick] (1.2, -1.8) rectangle (3.7,-2.6);
        \node[circle, inner sep=0pt, fill=white, minimum size = 5mm] at (3.7,-2.6) {B2};
    \end{tikzpicture}    
    \caption{Westward detail view of the temperature distribution on vertical slices through the model level volume, (a) using model level temperatures for the summer case, (b) using model level temperatures for the winter case, and (c) using vertical temperature gradients for the winter case.}
    \label{fig:case-temp-slices}
\end{figure*}
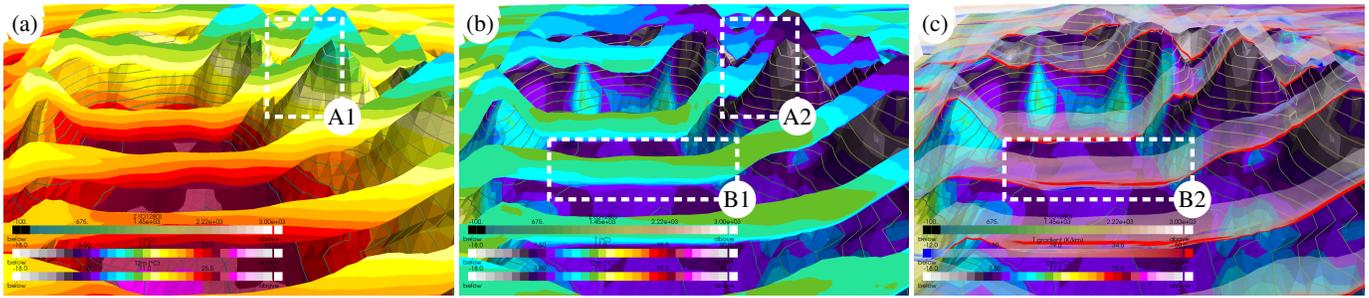

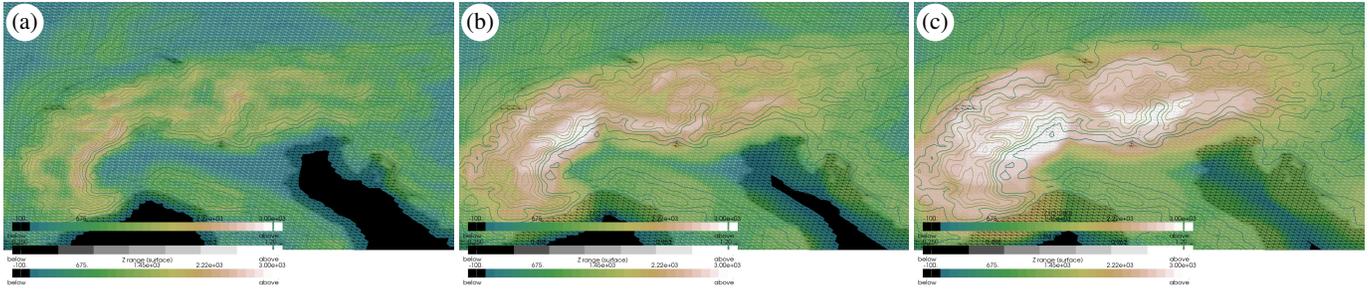
\begin{figure*}
    \centering
    \begin{tikzpicture}
        \node[inner sep=0pt, anchor=north west] (highres) at (0, 0) {
            \includegraphics[width=.33\textwidth]{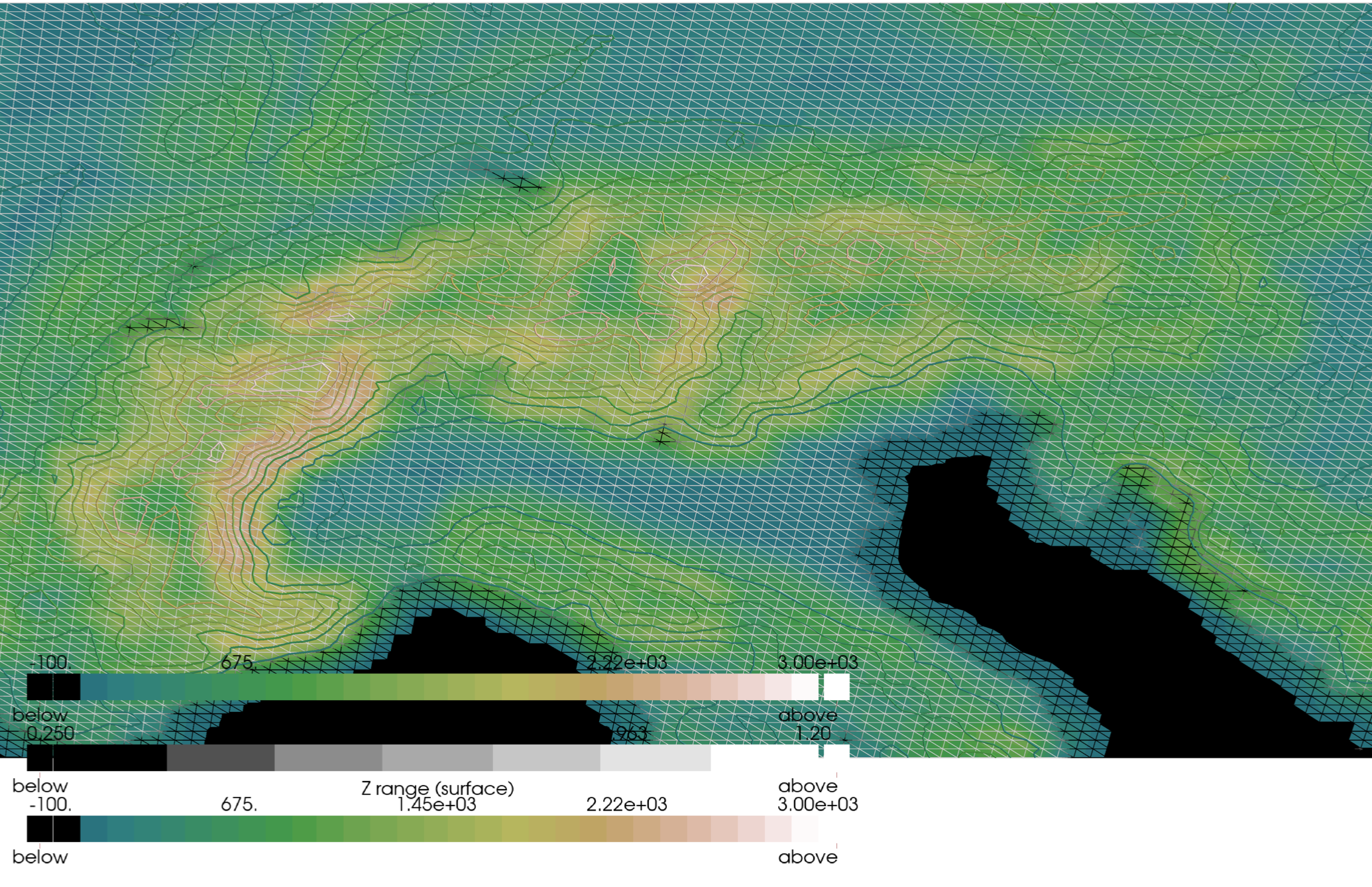}
        };
        \node[circle, inner sep=0pt, anchor=north west, fill=white, minimum size = 5mm] at (0.1, -0.1) {(a)};
    \end{tikzpicture}    
    \begin{tikzpicture}
        \node[inner sep=0pt, anchor=north west] (highres) at (0, 0) {
                \includegraphics[width=.33\textwidth]{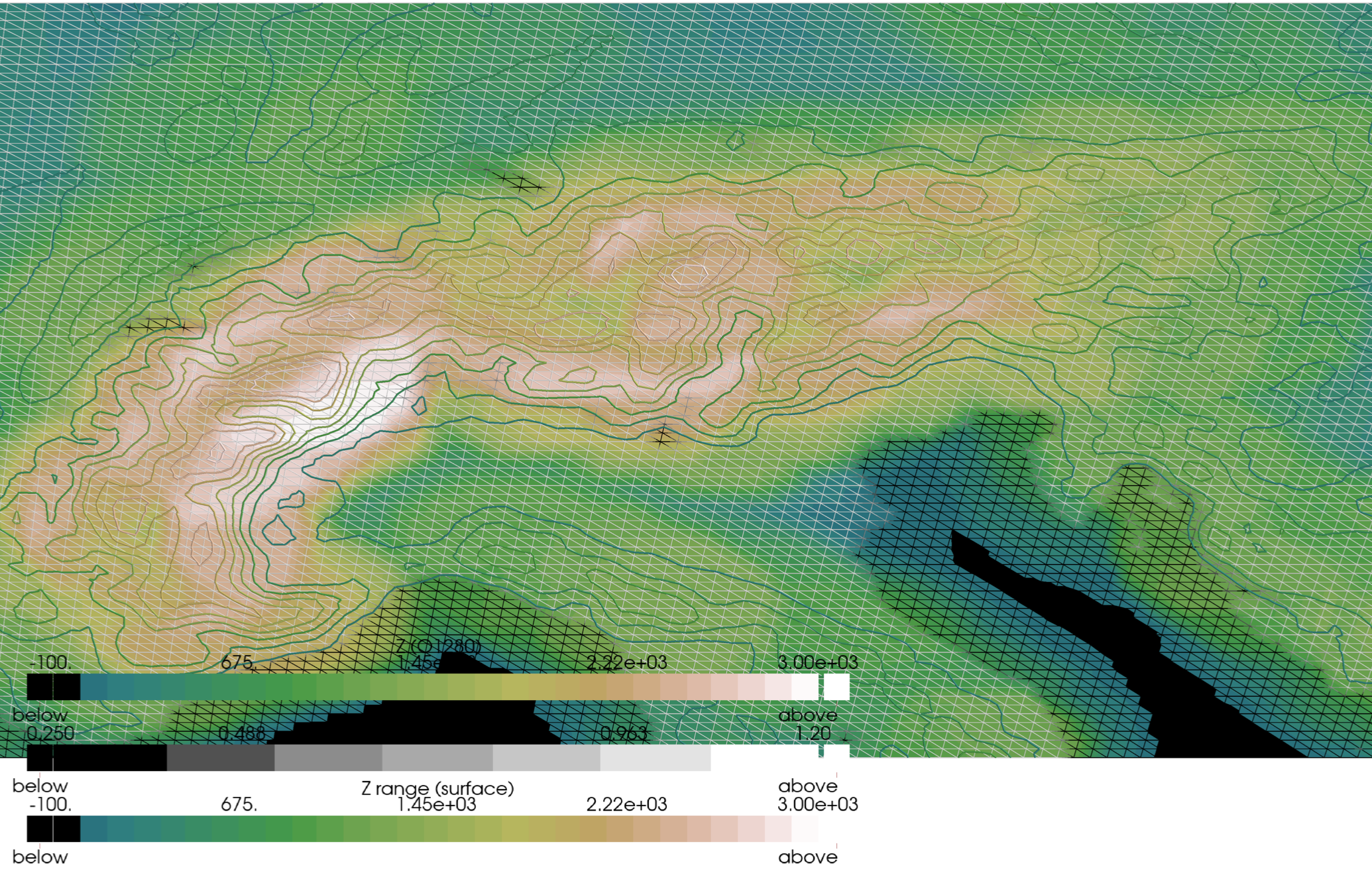}
        };
        \node[circle, inner sep=0pt, anchor=north west, fill=white, minimum size = 5mm] at (0.1, -0.1) {(b)};
    \end{tikzpicture}    
    \begin{tikzpicture}
        \node[inner sep=0pt, anchor=north west] (highres) at (0, 0) {
            \includegraphics[width=.33\textwidth]{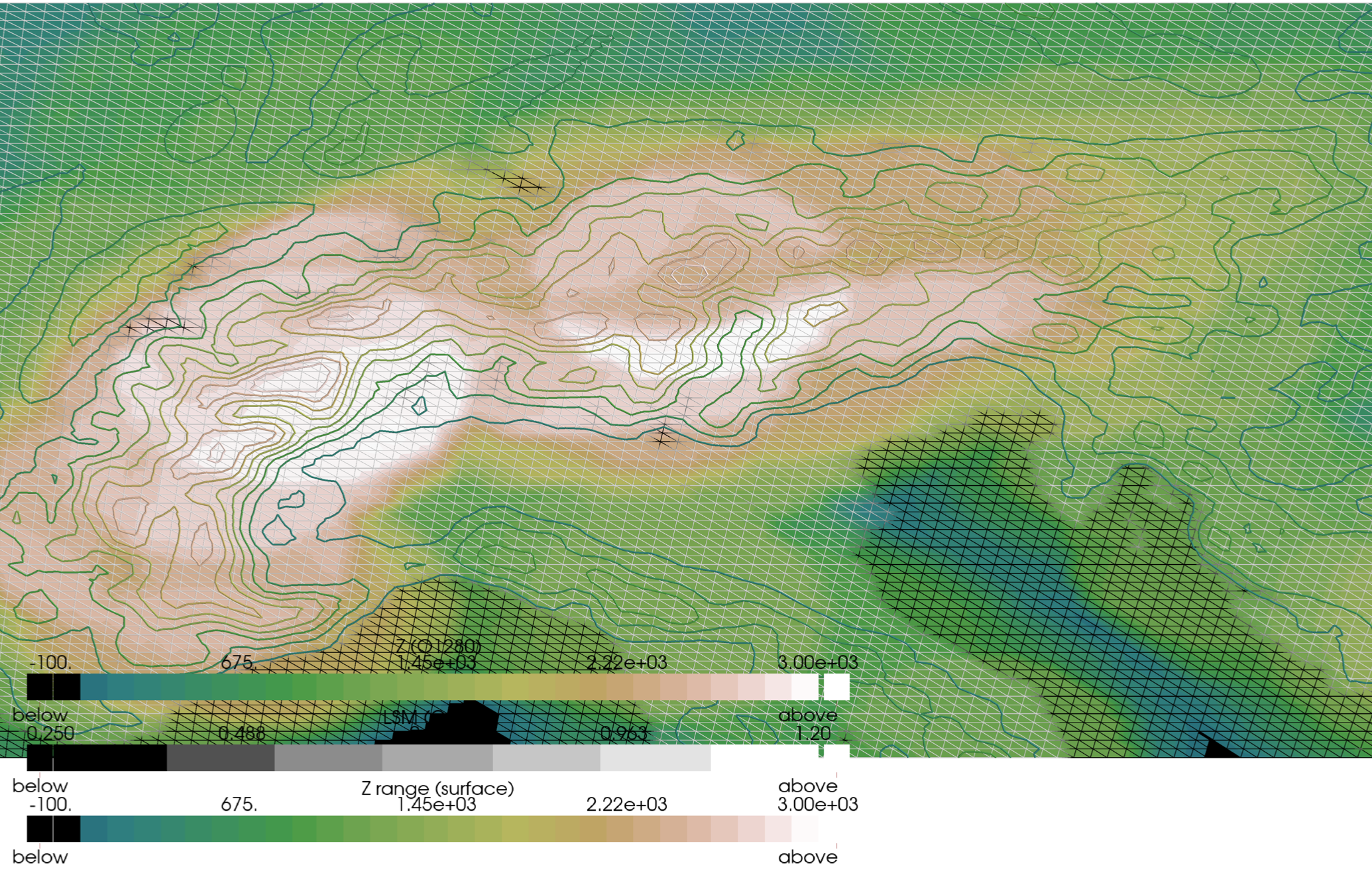}
        };
        \node[circle, inner sep=0pt, anchor=north west, fill=white, minimum size = 5mm] at (0.1, -0.1) {(c)};
    \end{tikzpicture}    
    \caption{Elevation range (surface color) and land-sea mask (grid lines) as seen by the lapse rate algorithm with radius settings (a) 30 km, (b) 60 km and (c) 90 km. Orography isocontours are shown for orientation.}
    \label{fig:lapserate-elevrange}
\end{figure*}

\begin{figure*}
    \centering
    \begin{tikzpicture}
        \node[inner sep=0pt, anchor=north west] (highres) at (0, 0) {
            \includegraphics[width=.49\textwidth]{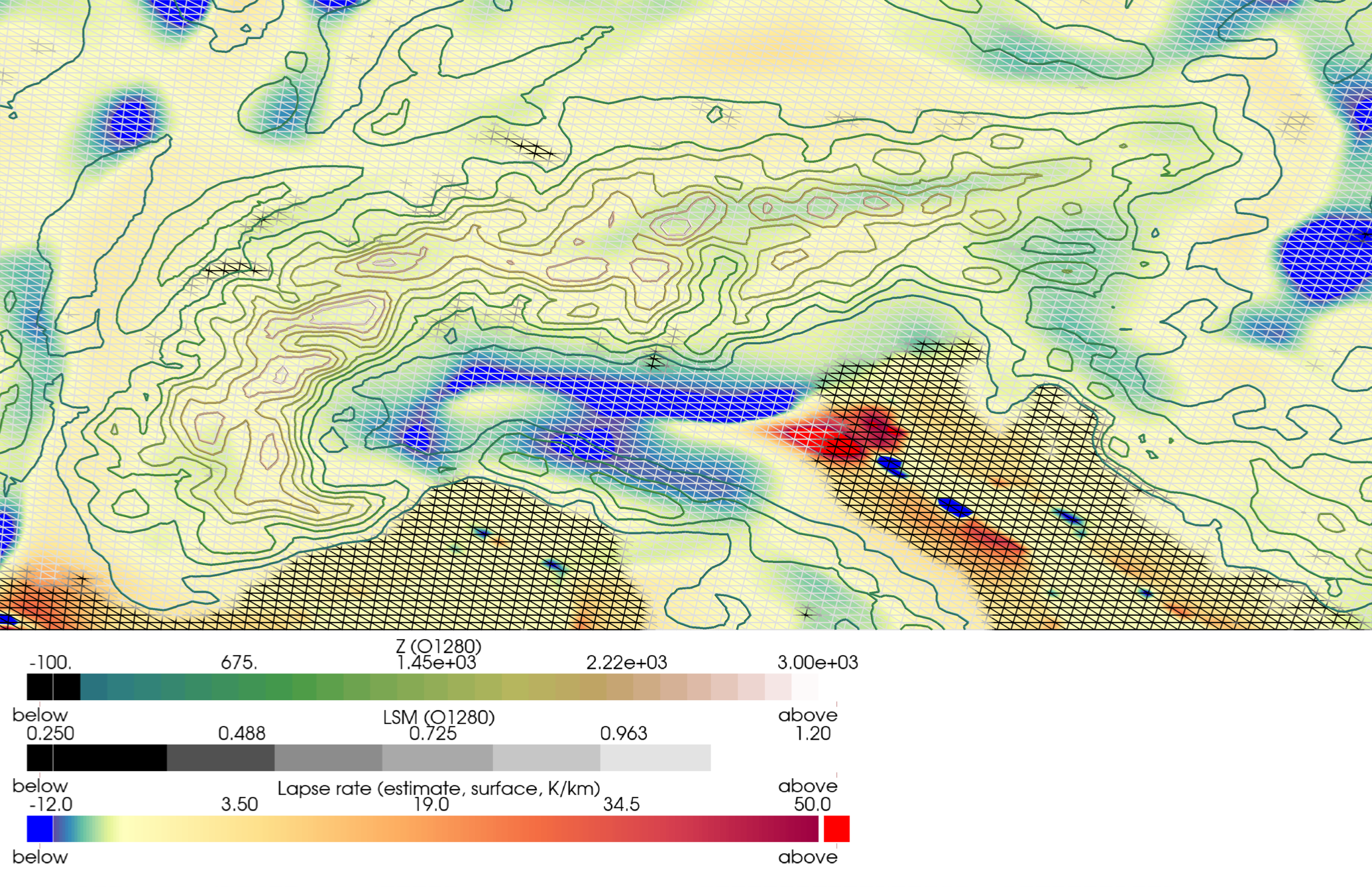}
        };
        \node[circle, inner sep=0pt, anchor=north west, fill=white, minimum size = 5mm] at (0.1, -0.1) {(a)};
    \end{tikzpicture}    
    \begin{tikzpicture}
        \node[inner sep=0pt, anchor=north west] (highres) at (0, 0) {
                \includegraphics[width=.49\textwidth]{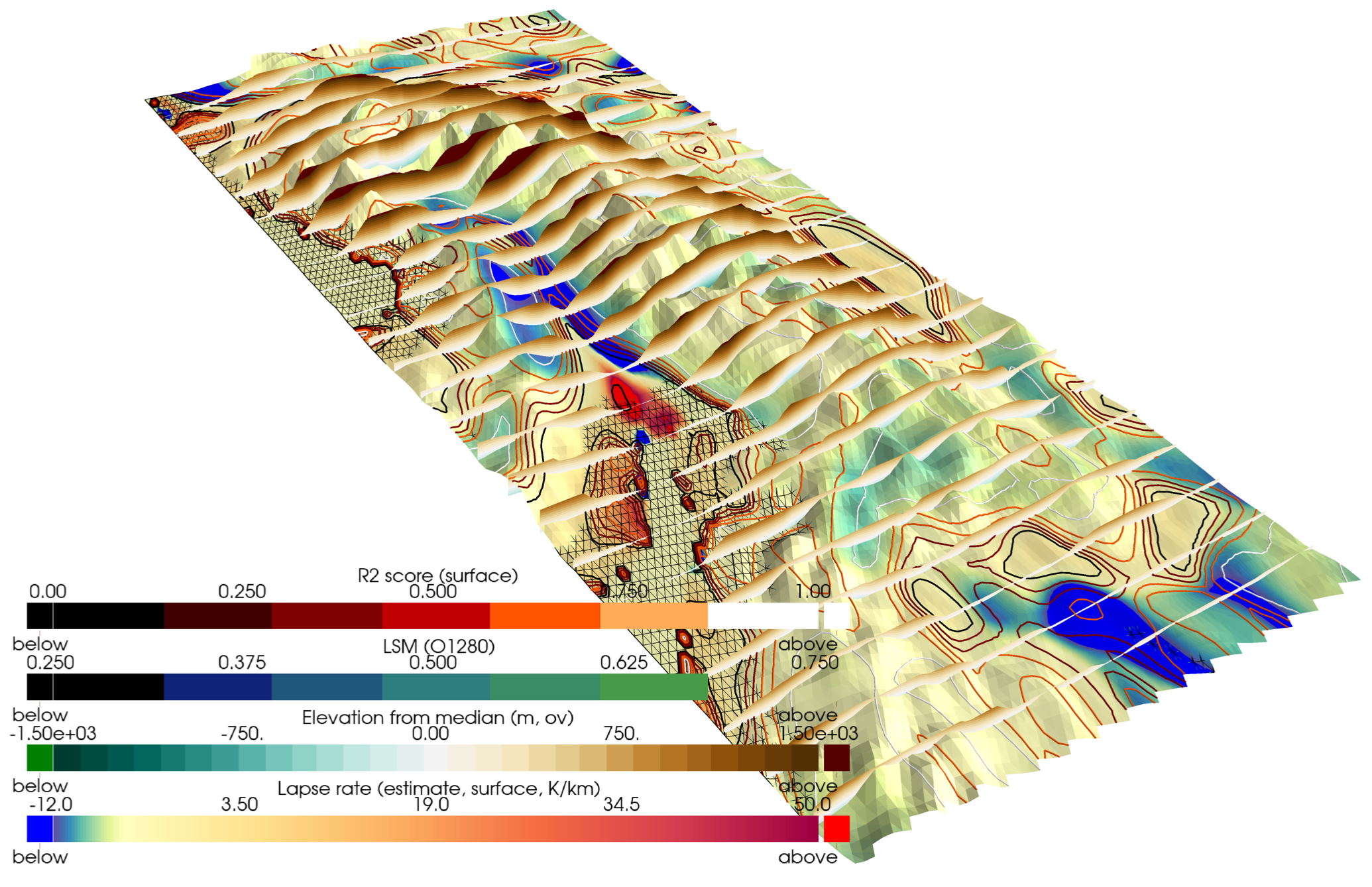}
        };
        \node[circle, inner sep=0pt, anchor=north west, fill=white, minimum size = 5mm] at (0.5, -0.1) {(b)};
    \end{tikzpicture}   
    \caption{Visualisation of summer-case lapse rates (a) in terrain context with orography contours and land-sea mask on grid lines, and (b) as a 3D view with clamping metrics terrain range and $R^2$ score displayed on elevation summary slices and isocontours, respectively, as well as land-sea mask on grid lines.}
    \label{fig:lapserate-metrics}
\end{figure*}

\begin{figure*}
    \centering
    \begin{tikzpicture}
        \node[inner sep=0pt, anchor=north west] (highres) at (0, 0) {
            \includegraphics[width=.49\textwidth]{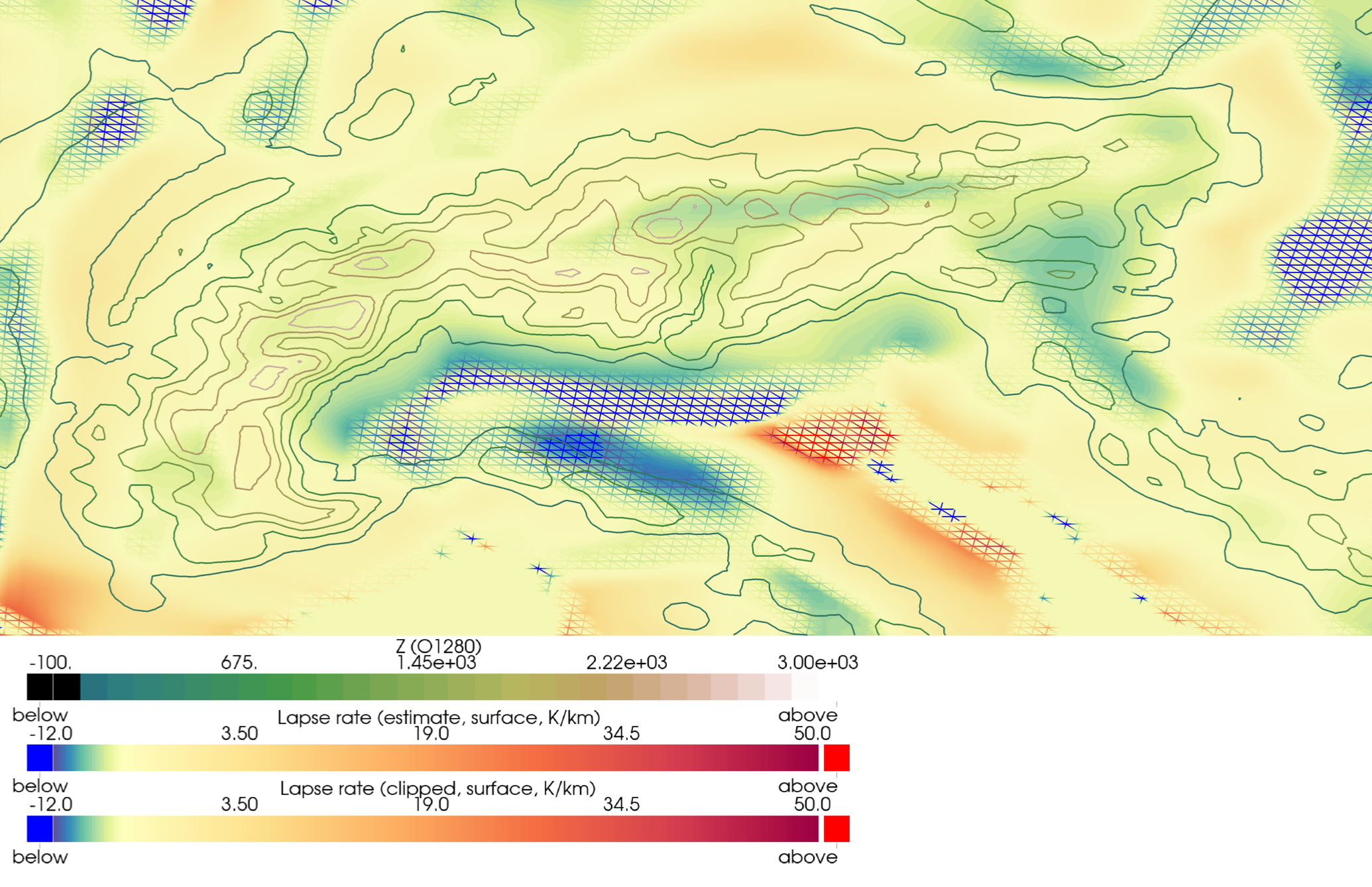}
        };
        \node[circle, inner sep=0pt, anchor=north west, fill=white, minimum size = 5mm] at (0.1, -0.1) {(a)};
    \end{tikzpicture}    
    \begin{tikzpicture}
        \node[inner sep=0pt, anchor=north west] (highres) at (0, 0) {
                \includegraphics[width=.49\textwidth]{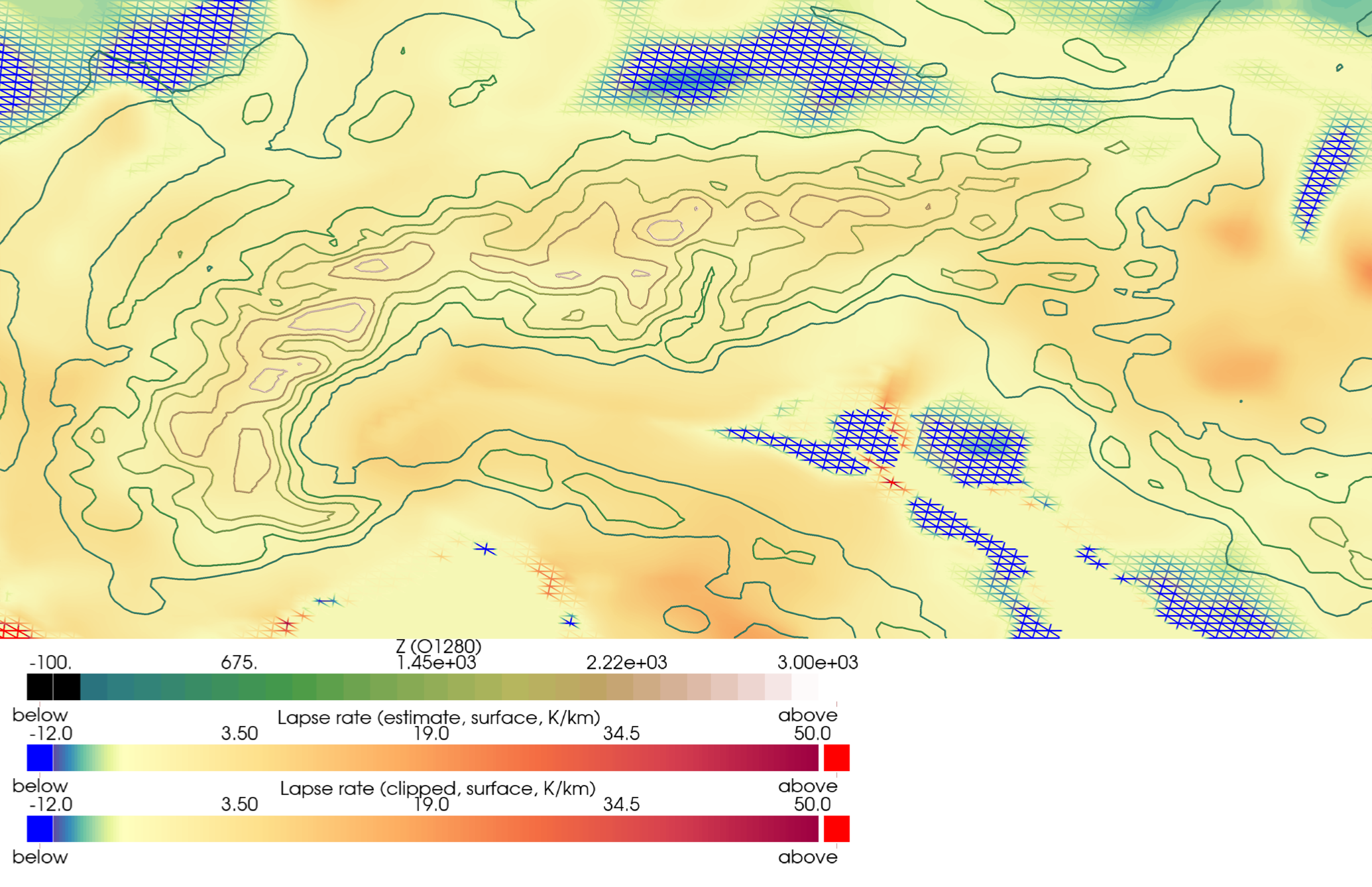}
        };
        \node[circle, inner sep=0pt, anchor=north west, fill=white, minimum size = 5mm] at (0.1, -0.1) {(b)};
    \end{tikzpicture}   
    \caption{Visualisation of lapse rates after clamping for (a) the summer case and (b) the winter case. Isocontours indicate the orography for orientation.}
    \label{fig:lapsrates-clipped}
\end{figure*}

\autoref{fig:case-temp-slices} (a) and (b) illustrate the distribution of model-level temperatures for the summer and the winter case using vertical slices through the temperature volume. The lighting of the slices has been turned off, and the density of elevation isocontours has been increased to achieve a visual contrast between the appearance of the terrain surface and the slice surfaces. For the user, this improves the possibility of accurately detecting the intersection between terrain and slice surfaces. Slicing is generally preferred over volume ray-casting due to the better run-time performance of the visualization, which enhances interactivity.

In the summer case, \autoref{fig:case-temp-slices} (a), the isolevels of the volumetric temperature are mostly flat, supporting the validity of a constant lapse rate scheme again. The isolines curve slightly upwards at the intersection between slices and terrain (see box A1). This may be interpreted as a sign of warm air masses rising on the terrain surface as superadiabats form due to ground heating by solar irradiation. Further clarity on this could be obtained using additional information on near-surface air movement but it is not of primary importance for the lapse rates and is, therefore, beyond the scope of this work.

In the winter case, \autoref{fig:case-temp-slices} (b), the isolevels near the surface indicate a significant temperature gradient between the lowest model level and the 2 m temperature. The gradient is manifested in a strong curvature of the isolines close to the surface (see box A2). To examine the vertical temperatures in more detail, \autoref{fig:case-temp-slices} (c) displays the same scene as \autoref{fig:case-temp-slices} (b), but with slice color encoding the vertical temperature gradient. The color mapping uses a dedicated diverging color scale, which respects the physical limits of plausible temperature gradients. The color scale is centered and $-6.5$ K/km, corresponding to the value of the default lapse rate of the temperature correction model, and is capped at $-12$ K/km due to the physical instability of air masses beyond that limit (10 K/km is the true standard for instability, using 12 K/km allows for some superadiabatic behavior at the 2 m level). A V-shaped opacity function is applied to minimize occlusions, which has a user-configured minimum opacity at the color scale center and higher opacity towards the extremes. By default, the visualization applies a linear slope for the opacity increase. However, to prune larger parts of the volume or increase the visualization's density, the user can switch to a polynomial opacity increase with exponents $>1$ or $<1$. \autoref{fig:case-temp-slices} (c) shows clearly the complexity of the volumetric temperature field. Close to the surface, temperature gradients of more than 50 K/km are observed. Inside the valley (boxes B1 and B2), multiple air layers with alternating gradient signs are stacked on top of each other, clearly invalidating the assumption of a constant lapse rate. Only at higher altitudes do the temperature gradients revert towards the regular value of $-6.5$ K/km.

\subsection{Visual analysis of the adaptive lapse rates}

To evaluate the quality and tune parameters of the proposed adaptive lapse rate scheme (\textbf{Q3}), we support the visual analysis of lapse rate scores and the associated clamping metrics $R^2$ score. For this, the user selects the required hyperparameter in the graphical user interface, obtains handles to view the lapse rate estimates, and applies the clamping as a postprocessing step. 

Visualizing the data is challenging since multiple aspects of the local model and terrain data determine the lapse rate. The primary input variables are the orography and the land-sea mask, which determines whether a grid vertex is used as a valid sample location. The threshold for evaluating the land-sea mask can be set interactively, but a value of 0.5 is a good default as this is implicit in the model formulation. 

The radius parameter is more critical and is explored in \autoref{fig:lapserate-elevrange}. The figures display the range of elevation values used in computing the lapse rate estimate. Notably, the range size changes discontinuously, especially for radii 60 km and 90 km, when grid vertices with extreme elevation enter or fall out of the radius neighborhood (circle patterns in \autoref{fig:lapserate-elevrange} (b) and (c)). Using a hard cutoff radius also leads to a discontinuity in the lapse rate estimates and motivates the use of a distance-based Gaussian weighting scheme. This way, more weight is put on vertices closer to the reference location during the lapse rate estimation. Empirically, a cutoff radius of 60 km in combination with a Gaussian weight scale of 30 km yields the best balance of stability of the estimates and feature resolution.

\autoref{fig:lapserate-metrics} displays lapse rate estimates for the summer case using these parameters. It appears that the estimator yields extreme lapse rate estimates. In \autoref{fig:lapserate-metrics} (a), the estimates are displayed in the land-sea mask and orography context. It is seen that many of the extreme cases arise in the transition region between land and sea, where lapse rate estimation tends to be difficult, and adaptive lapse rate estimation using temperature and elevation samples is ill-defined due to the lack of orographic variation. Therefore, the scheme reverts to the default lapse rate for sea-site locations. One should also note that "extreme" lapse rates diagnosed in topographically almost-flat areas are inconsequential for 2 m temperature reconstruction, as the elevation difference multipliers are so small. 

In \autoref{fig:lapserate-metrics} (b), the volumetric display includes additional information about the lapse rate reliability metric $R^2$. Values of $R^2$ are shown as isocontours on the terrain and are color-coded by their respective value. For additional context, an elevation range summary is added as vertical slices, indicating the elevation values range seen by the lapse rate estimator (cf.\ \autoref{fig:lapserate-elevrange}). It can be seen that large values of $R^2$ may occur both in regions with large elevation ranges and in areas with minimal elevation ranges. Especially in the latter regions, the resulting lapse rate estimates exhibit extremes, which appear unreasonably large and are more likely a numerical sampling artifact rather than evidence in favor of a positive or negative lapse rate different from the default. Like sea-site locations, such stations are handled by reverting to the default lapse rate. 

Based on the user interface, the clamping of the lapse rates can be optimized. \autoref{fig:lapsrates-clipped} displays maps with the clamped lapse rate encoded in the surface color and the raw lapse rate in the color of the mesh parameters. A difference in overall shade can be perceived, indicating more negative values for the summer case (blue shades) and more positive lapse rates (red shades) for the winter case. 

\subsection{Quality of the lapse rate estimator\label{sec:quality}}

Despite the possibility to visualize predictions and lapse rates, decisions on the parameterization of the lapse rate scheme have to be put on a statistical footing (\textbf{Q4}). The observation dataset is used for this purpose as follows. The observations are first grouped by the station that generated the data. Then, from the pool of ca.\ 14500 available stations, 20\% are selected for parameter tuning, and the remaining are kept for testing. The station groups are sorted by the amount of available observations and split into consecutive groups of 5 stations each. From each group, one station is selected randomly, and the observations of this station are moved to the training dataset. This procedure yields training and test datasets with a similar average number of observations per station and minimizes information overlap between training and test data. Note that the fraction of training data is chosen low compared to other statistical estimation tasks. This is done because the low parameter count of the models justifies a reduction of training data. More data is available for verification.

Parameter tuning involves mainly the threshold settings of the clamping functions. For this, lapse rates and $R^2$ scores are computed for all available observations and are divided into ten groups according to the value of $R^2$. For each group, a grid search is performed to identify the set of fixed upper and lower bounds on the lapse rate, for which the prediction accuracy is optimized. The parameterization of the clamping functions \autoref{eq:clamps} is obtained by comparing and balancing bound parameters of all $R^2$ groups.

For quantitative testing, lapse rates, $R^2$ scores, and predictions are computed for observations from the evaluation dataset. The stations are grouped into ten bins according to the value of $R^2$ and are classified further concerning their elevation difference against the model terrain. As in \autoref{sec:visana}, stations more than 50 m below the model terrain are classified as concave (valley) stations, whereas stations more than 50~m above the terrain are called concave (mountain) sites. Stations in between are called neutral. 

\begin{figure}[t]
    \centering
    \includegraphics[width=.45\textwidth]{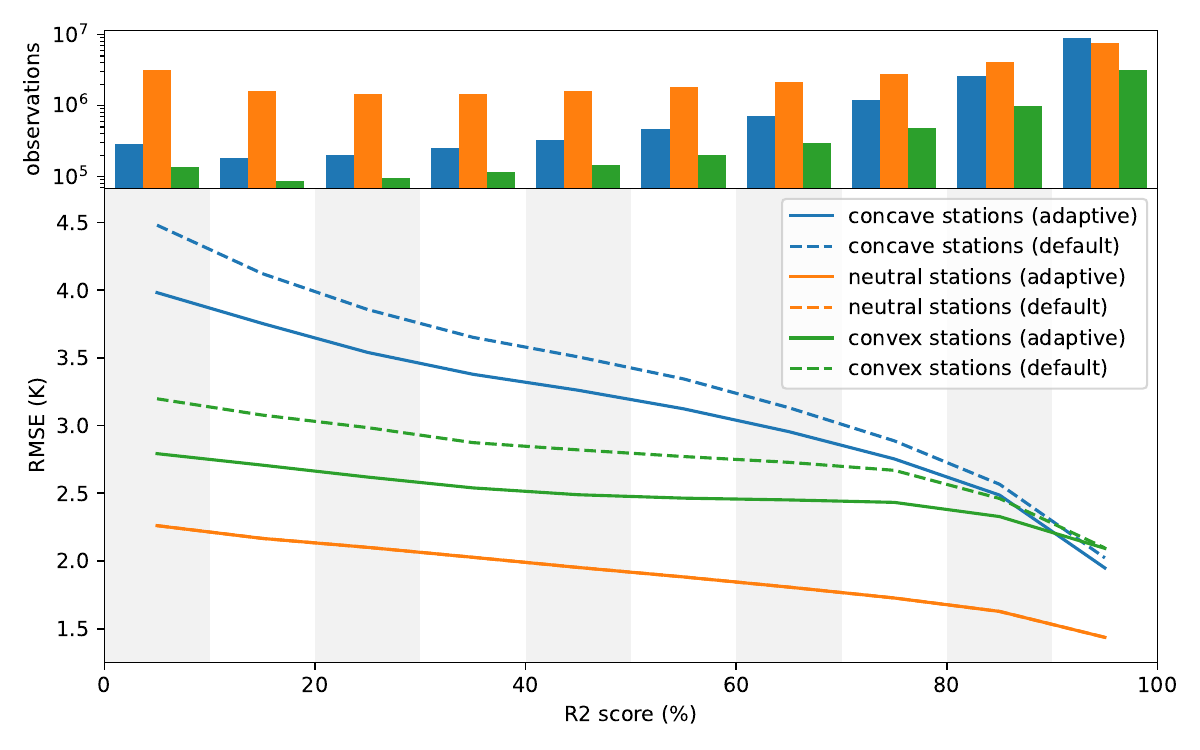}
    \caption{RMSE prediction error of the adaptive lapse rate scheme compared to the default scheme. Observations are binned according to the observed $R^2$ score. Histograms indicate the number of observations falling in each bin for valley-site, mountain-site, and neutral stations. In difficult local weather conditions (low $R^2$), predictions are improved by 10-20\% for both valley and mountain stations.}
    \label{fig:rmse}
\end{figure}

\autoref{fig:rmse} shows the predictions' root mean squared error (RMSE) against the temperature observations. For both valley and mountain stations, the adaptive lapse rate scheme improves the prediction accuracy by between 10\% and 20\%. The histograms indicate the amount of observations falling into each bin during the evaluation.

\begin{figure}
    \centering
    \includegraphics[width=.45\textwidth]{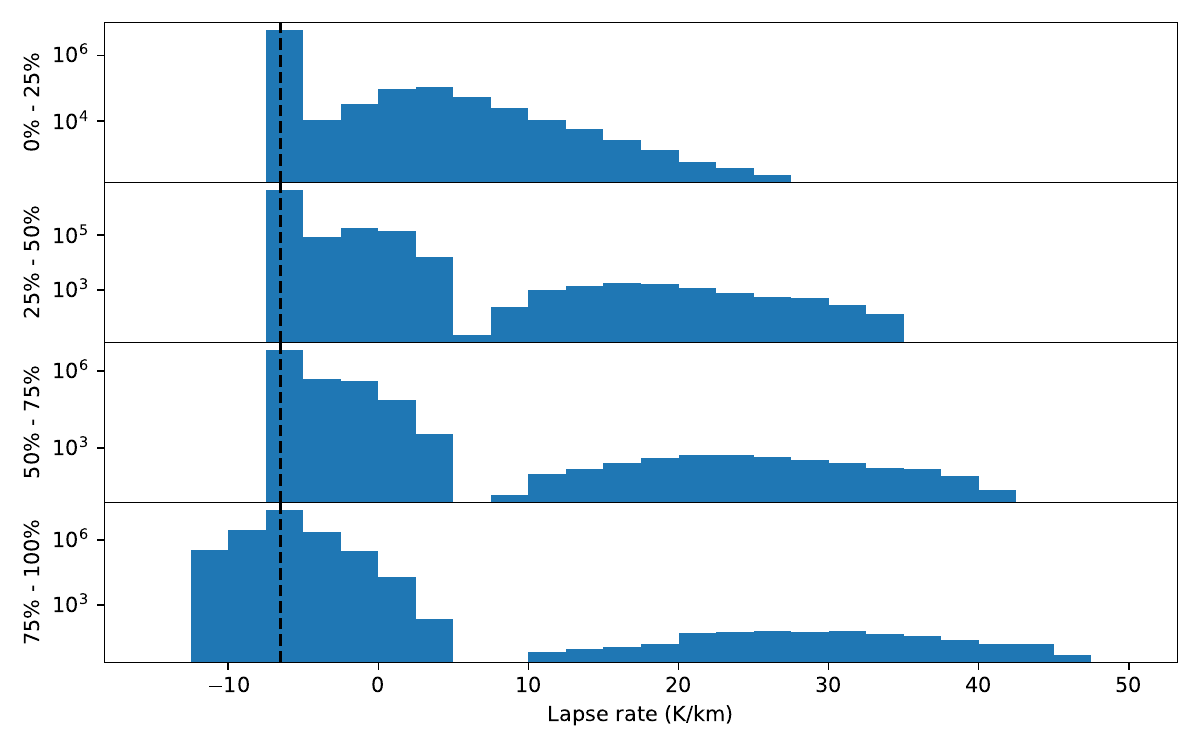}
    \caption{Histograms of adaptive lapse rates after clamping for different ranges of the observed $R^2$ score. Ranges are given on the vertical axes. The black dashed line indicates the default value of $-6.5$ K/km.}
    \label{fig:lrhist}
\end{figure}

\autoref{fig:lrhist} visualizes the distribution of estimated lapse rates for different levels of the observed $R^2$ score. The lapse rates are shown after clamping, which explains why no lapse rates with values below $-6.5$ K/km are observed for the score groups below 75\%. 
However, the mode of the distribution persists even in the most determined category. This is reassuring since the models often suggest physically sensible lapse rates. It is seen that with increasing determination coefficient, the lapse rates develop an increasingly bimodal distribution with growing variance. No lapse rates are observed between 5 and 10 K/km among the most determined models. The histograms demonstrate that the lapse rate scheme identifies weather situations in which the local lapse rates deviate from the default. 

\section{Discussion and Conclusion}

We have applied existing and developed new topographic visualization techniques to assess the quality of lapse rate schemes in the context of low- and high-res orography. Through these techniques, relations between lapse rate quality, orographic features, and seasonal conditions could be revealed. The visualizations have helped to spot specific relationships that have been considered in developing an improved adaptive lapse rate scheme. 

\autoref{fig:rmse} shows that our adaptive lapse rate scheme, on average, improves 2 m temperature RMSE in concave and convex locations. Such improvements are all the more striking given the relatively high frequency of cases where the dynamic lapse rate is similar to the standard value and errors are the same. Concave sites exhibit larger errors than concave in both lapse rate schemes, perhaps because of the particular challenge of cold air pooling (which attracted an additional postprocessing step in \cite{sheridan_2}). Such errors fall into the class of situation-dependant systematic model biases, which can be an Achilles heel for attempts to reduce downscaling errors \cite{sheridan_main}. Though not included directly in this study, a further global postprocessing step could be included in a future operational incarnation of our dynamical lapse rate adjustment. This is, for instance, to use the ECMWF postprocessing framework ecPoint \cite{ecpoint} to apply, in tandem, situation-dependant grid-scale bias correction to raw model 2 m temperatures (Figure 3 (a), (b), (d) in \cite{highlander} is an example).

An interesting question for future work is the role of outliers in the local environments. In the present study, lapse rate estimation is performed with linear regression models. Despite achieving considerable improvements in prediction skill, Figures \ref{fig:lapserate-metrics} and \ref{fig:lapsrates-clipped} show that there exist weather situations in which the lapse rate scheme suggests extreme lapse rates beyond physical plausibility. The issues become more visible the smaller the environment radius is selected.
Clamping has been identified as a countermeasure but may lead to information loss at the cost of reduced prediction accuracy. We believe that robust regression methods (which are more stable against outlying temperature and elevation samples than standard linear regression) may help to improve the accuracy further. As such models have their own intricacies and limitations, this aspect is left for future work.

A statistically robust extension of the proposed scheme would also be suitable for downscaling other meteorological variables. While we have focused on 2 m temperature in this study, rainfall is an equally important variable for forecast users in certain applications. Precipitation, too, can have a strong but variable topographic dependence. Therefore, a possible extension of this work would be investigating a similar dynamical lapse-rate scheme for rainfall downscaling, wherein "lapse rate" would reference rainfall rate or rainfall totals.

User motivation for the improved lapse rate scheme derives from the most-used graphical product out of the many produced operationally by ECMWF, namely meteograms. These display the range of possible forecast outcomes in the upcoming 10-15 days for a handful of key surface weather parameters, including 2 m temperature. The user selects a site, and then, via the fixed lapse rate assumption, adjustments are automatically made to deliver the meteogram 2 m temperatures for them. The uncertainty range bounds are derived from the use of multiple (ensemble) forecast realizations. Due to the computational simplicity of the adaptive approach, the novel scheme is perfectly portable to forecast ensembles. We envisage that an operational implementation of our approach would replace the fixed lapse rate assumption in this processing chain with a situation-dependent variable lapse rate.

The used visualizations show that while 3D representations often help and are even necessary to convey the relevant information, at the same time, they can hinder effective information communication due to occlusions and visual clutter when overloaded with too many additional, yet functional, visual mappings. Our analysis shows that 2D maps are still indispensable for topographic data visualization, especially in operational use. In the future, we intend to develop dedicated 2D map views to effectively convey the many aspects that need to be considered in operational forecast products, and we will equip them with linked 3D views to support specific investigations and/or representations. 

Given that numerical models now represent the vertical structure of the atmosphere in great detail, one could imagine that extraction from a model profile would suffice for higher elevations. For relatively isolated peaks, this can work, but within the terrain, near-surface temperatures can be decoupled from the free atmosphere due, for example, to the sometimes strong influence of surface fluxes (see \cite{sheridan_2}). The 3D visualizations developed in this work will be helpful in investigating such aspects since they allow for a comprehensive view of different atmospheric conditions and their interplay with orography.

\acknowledgments{%
    The authors acknowledge funding by the Munich Center for Machine Learning (MCML) initiated by the Federal Ministry of Education and Research and the State of Bavaria.
}

\bibliographystyle{abbrv-doi-hyperref}

\bibliography{template,BibPromo4}

\appendix 
\end{document}